\documentclass[preprint,tightenlines,eqsecnum,aps,epsfig]{revtex4}
\usepackage{graphicx}
\usepackage{epstopdf}
\usepackage{epsfig}
\usepackage{pdfsync}
\def\beq{\begin{equation}}
\def\eeq{\end{equation}}
\begin{document}
\title{
Increase with energy of parton transverse momenta in the
fragmentation region in  DIS and related phenomena. }
\author{ B. Blok\email{Email :blok@physics.technion.ac.il} }
\affiliation{Department of Physics, Technion---Israel Institute of Technology, 32000 Haifa, Israel}
\author{ L. Frankfurt\email{E-mail: frankfur@tauphy.tau.ac.il} }
\affiliation{School of Physics and Astronomy, Raymond and Beverly
Sackler Faculty of Exact Sciences, Tel Aviv University, 69978 Tel
Aviv, Israel}
\author{ M.Strikman\email{\E-mail: strikman@phys.psu.edu}}
\affiliation{Physics Department, Penn State University, University Park, PA, USA}
\thispagestyle{empty}
\begin{abstract}
 The dipole and the DGLAP approximations are combined with the
 $k_t$ factorization theorem to demonstrate the fundamental
 property of pQCD: smaller is the size of the colorless quark-gluon
 configurations in the fragmentation region, more rapid is the
 increase of its interaction with the target as a function of
 energy. First, we consider two closely related properties of the wave function of the
 projectile  (i)
   the transverse momenta of the quark(antiquark) within the $q\bar{q}$ pair, produced
   in the fragmentation region by the strongly virtual photon, increase with the decrease
   of x for fixed $Q^2$, (ii) increase of the relative
contribution of pQCD  to the structure functions
  as compared to soft QCD contribution at central impact parameters
due to a rapid increase with energy of the cross section of
interaction of small dipoles. Practical consequences of these
effects are presented for the cases of the
 cross sections of DIS and  DVCS. We
  predict that the ratio of DVCS to DIS amplitudes should very
 slowly approach
one from above   at very large collision energies. Second, we
study a closely related phenomenon of the
 increase of the
 transverse momenta with the energy of the characteristic transverse
 momenta
 of the gluon/quark configurations responsible for the transition to
 the black disk regime.
  We discuss
 the impact of this phenomenona on the slowing of the dependence on
 the initial energy of the coherence length. We demonstrate that a
 rapid projectile has the biconcave shape, which is different from
 the expectations of the preQCD parton model where a fast hadron
 has a pancake shape. We show that the increase of the transverse
 momenta leads to a new
expression
 for the total cross section of a
 DIS scattering
 at very large energies, relevant to LHeC and LHC. We
 discuss the impact of the discovered phenomena on the hard
 processes in pp collisions, and on the dominance of different
 phases of chiral and conformal symmetries in the central and
 peripheral pp, pA, and AA collisions.
\end{abstract}
\maketitle \setcounter{page}{1}
\section{Introduction.}
\par
The leading order dipole approximation for the high energy
processes in QCD developed in Ref. \cite{BFGMS}, cf.
 also Refs. \cite{AFS,FMS,BBFS,Muellereikonal,CFS},
 is the generalization of the pQCD  improved
parton model  to the target rest frame description.
 The leading order dipole approximation provides the solution of
the equations of QCD in the kinematics of fixed and small, but not
too small $x=Q^2/\nu$ and not very small $Q^2$.
\par The main aim of the current paper is the study of the
properties of the DIS processes in the fragmentation region. In
particular, we study two closely related properties of the
wave function of the projectile. \par  First,  we consider two
 properties of the perturbative QCD (pQCD) wave function of the
 projectile:  (i) we find the increase with energy (1/x) of
   the transverse momenta ($k_{\rm t}$) of the quark(antiquark) within the $q\bar{q}$ pair produced
   in the fragmentation region by the strongly virtual photon, (ii) we see the increase of the relative
contribution of pQCD  to the structure functions
  as compared to the soft QCD contribution at central impact parameters
due to a rapid increase with energy of the cross section of
interaction of small dipoles. We show that at sufficiently high energies
the transverse  momenta $k_t$  become
larger than $Q^2/4$.
 As a result  a new pQCD regime different from the conventional
DGLAP \cite{DGLAP}, appears at these energies. The reason is that
the ordering in transverse momenta that led to large ${\rm log}
(Q^2/\Lambda^2)$ is not valid any more.
 Practical consequences of these
effects are presented for the cases of the
 cross sections of DIS and  DVCS. We
  predict that the ratio of DVCS to DIS amplitudes should decrease with energy and very
 slowly approach
one from above   at very large collision energies. In addition the ratio $\sigma_L/\sigma_t\propto Q^2/(4k^2_t)$
will also slowly decrease with energy. Technically these effects follow from the
more rapid increase with the energy of the pQCD interaction
 for smaller dipole and the $k_t$ factorization theorem \cite{CH,CE}.

\par Second, we
study a closely related phenomenon of the
 increase with the energy of the characteristic
 transverse momenta
 of the gluon/quark configurations in the  wave function of the projectile
 responsible for the transition to
 the black disk/Froissart  (BDR) regime.
  We discuss
 the impact of this phenomena on the slowing of the rate of increase  of the coherence length as
 a function of  energy.
The most striking  implication of our analysis is the change of  the space
 structure of the wave packet describing a rapid hadron. In the
classical multiperipheral picture of Gribov a hadron has
a shape of a pancake of the longitudinal size $1/\mu$ (where $\mu $ is the scale of soft QCD) which does
 not depend on the incident energy \cite{Gribovspace-time}. On the contrary, we find
  the biconcave shape for the rapid hadron in pQCD with the minimal
longitudinal length (that corresponds to small impact
 parameter $b$) decreasing with increase of energy and being smaller for the nuclei than for the nucleons.
\par The known rate of increase of the black disk regime
momenta leads to the generalization of the old Gribov formula for
the total DIS cross section in the black disk regime. We are able to
establish an overall  coefficient in the cross section dependence
on the $ \log^3(1/x)$ in the black disk regime. We
determine the relative contributions of pQCD and black disk regime
into the total small dipole DIS cross section at high energies.
This new
expression
 for the total cross section of a
 DIS scattering
 at very large energies  is relevant to LHeC and LHC.
\par It is worth  noting that the effect of the increase of the transverse
momenta in the fragmentation region in pQCD found in this paper
  is very different from the seemingly similar
effect
 found in the leading $\alpha_s\log(x_0/x)$ BFKL approximation\cite{BFKL}: for
 the central rapidity kinematics $\log^2(k_{t}^2/k_{t 0}^2)
\propto \log(s/s_0)$ . The latter is the property of the radiation
within a ladder, i.e. of a diffusion in the space of the transverse
momenta \cite{BFKL}. Indeed, it has been known for some time
already that if we look at characteristic transverse momenta in a
rung with a fixed number N in BFKL ladder, than in the multiregge
kinematics the transverse momenta do not depend on energy. This
fact follows from the derivation of Lipatov diffusion equation,
where $\log (k^2_t/k^2_{t 0}) \propto  N$-the number of the rung
under study. The Lipatov diffusion arises since a number of rungs
$N$ in the ladder increases with the rapidity Y. An alternative
proof that the transverse momenta do not rise in multiregge
kinematics with a fixed number of rungs has been given in Ref.
\cite{Sherman}. On the other hand the property we are dealing here
with is the value of the transverse momenta in the wave function
of the projectile.
\par The rapid increase with the energy of the characteristic transverse
scales in the fragmentation region in the black disk regime has
been found first in Refs. \cite{FGMS,FSW,Rogers1,Rogers2}.
 The prediction  of the increase with energy
 of the transverse momenta in the impact factor , in
 the kinematical domain where methods  of the pQCD are still applicable can  be
 considered as
  a precursor of the black disk regime indicating the possibility of the smooth
  matching between the pQCD and black disk
  regimes.
 \par The characteristic
feature of the LO dipole solution is the approximate Bjorken
scaling for the structure functions of DIS, i.e. the two
dimensional conformal invariance for the moments of the structure
functions. In this approximation as well as within the leading
$\log(x_0/x)$ approximation, the transverse momenta of quarks
within the dipole produced by the local
 electroweak current are restricted by the virtuality of the external field:
\begin{equation}
\Lambda^2 \le k_t^2 \le Q^2/4.
\end{equation}
Here $\Lambda\equiv \Lambda_{QCD}=300 $ Mev is a QCD scale. It
follows from the QCD factorization theorem proved in Refs.
\cite{CH,CE} that within this kinematical range the smaller
transverse size $d$ of the configuration
 (the transverse distance between the constituents of the dipole) corresponds to a
 more rapid increase of its interaction with the collision energy:
\begin{equation}
\sigma= \alpha_s(c/d^2)F^2\frac{\pi^2}{4} d^2
xG(x,c/d^2),\label{1.3}
\end{equation}
here $F^2=4/3$ or $9/4$ depends whether the dipole consists of
color triplet or color octet constituents and is a SU(3) Casimir
operator,  $G$ is an integrated gluon distribution function and
$c$ is a parameter $c=4\div 9$. It is well known in the DGLAP
approximation that the structure function $G(x,Q^2)$ increases
more rapidly with $1/x$ at larger $Q^2$. This property agrees well
with the recent HERA data.

\par
Our main result  is that the median transverse momenta $k^2_t$ and
invariant masses of the leading $q\bar q$ pair in the
fragmentation region in the pQCD regime grow as
\begin{eqnarray}
k^2_t&\sim& a(Q^2)/(x_0/x)^{\lambda (Q^2)},\nonumber\\[10pt]
M^2&\sim& b(Q^2)/(x_0/x)^{\lambda_{\rm M} (Q^2)}.\nonumber\\[10pt]
\label{main}
\end{eqnarray}
Here $k^2_t$ and $M^2$ are the median squared transverse momentum
and invariant mass of the quark-antiquark pair in the
fragmentation region.  (The median  means that  the configurations
with the momentum/masses less than the median one contribute half
of the total crosssection). The exponential factors
 $\lambda$ and $\lambda_{\rm M}$ are both  approximately
 $\sim 0.1$.  These  factors are
  weakly dependent on the external virtuality $Q^2$. Their
exact values  also depend on the details of the process, i.e.
whether we consider the DIS process
 with longitudinal or
transverse photons, as well as on the model and approximation
used. This dependence, however,
turns out to be very mild. The exact form of $\lambda (Q^2)$, and $\lambda_{\rm M}
(Q^2)$ is given below.
\par The function $b(Q^2)\sim a(Q^2)/4$ for longitudinal photons
due to relation \beq
M^2=\frac{k^2_t+m^2_q}{z(1-z)},\label{con}\eeq  and the dominance
of the $z=1/2$ configurations. For transverse photons however
$a(Q^2)\ll b(Q^2)$ for realistic energies due a dominant
contributions of configurations with  $z\sim 0,1$ in the total
crosssections, i.e. $M^2\gg 4k^2_t$  due to the relation
\ref{con} ($z$ is the fraction of the total momentum of the dipole
carried by one of its constituents). The characteristic invariant
masses of transverse $q\bar q$ pair are of the order $Q^2$ at
realistic energies.

\par
We carry out the calculations in a wide region of a phase space:
\beq 10^{-7}<x<10^{-2},\,\,\,\, 3\,\, {\rm GeV}^2<Q^2<100\,\, {\rm
GeV}^2. \label{region} \eeq
We also obtain the increase of the median transverse momenta and
invariant masses for the jet distributions for
 both longitudinal and transverse photons.

\par
Although the increase of the transverse momenta in the
fragmentation region was obtained in this paper in the LO
approximation of pQCD it should be valid in the NLO as well.
Indeed, it is known that the NLO radiative corrections to the
impact factor are small \cite{Fadin,SP}. There are indeed large
radiative corrections to the ladder however these corrections are
already included in the LO  CTEQ functions, that were matched to
phenomenological data for the experimentally studied region of
$Q^2,x$.
\par
Let us note that the transverse momenta for the case of the
transverse photons were calculated in the framework of the
perturbative QCD. In order to ensure separation from  soft physics
(the AJM model), we calculated the average parton momenta with the
cutoff $k^2_t\sim 0.35$ GeV$^2$. The relative contribution of the
AJ  quickly decreases with energy, although it  is still considerable at
HERA energies and  $Q^2$ few GeV$^2$.

\par
Finally,   we discuss the possible applications of our results to
pp, pA collisions at the LHC. We explain that characteristic
phenomena   at central impact parameters are the blackening of the
interactions, the increase with the energy of the parton momenta,
and suppression of soft QCD contributions. At sufficiently large
energies the collisions at central impact parameters will be
dominated by a phase with unbroken chiral and conformal symmetries
. On the contrary, the peripheral collisions will be dominated by
a  phase with the  spontaneously broken symmetries.
 Applicability of this picture to QCD physics
at LHC will be subject of further research.
\par The paper is organized in the following way.
First, in section 2 we derive the spectral representation for the
total crosssection of the DIS. The derivation is done within the
framework of dipole model and $k_t$ factorization and generalizes
 LO DGLAP \cite{DGLAP} and BFKL \cite{BFKL} approximations. Next in section 3
  we derive eq. \ref{main} in double log approximation. This
derivation has an advantage that the increase of the transverse
momenta can be seen analytically, in a model independent way.
Next, in section 4 we derive eq. \ref{main} in LO approximation
using CTEQ5 and CTEQ6 structure functions. We find the kinematical
boundary of the new pQCD regime. We also discuss the influence of
the soft (AJ) contributions on our results. In section 5 we
applied our results to DCVS processes and study the dependence of
the cross-section of these processes on $Q^2$ and energy $x_B$. In
section 6 we derive the generalisation of Gribov formulae-- the
formula for total DIS cross-section in the black disk region, that
takes into account the rapid rise of the transverse momenta in the
fragmentation region in the black disk regime. In section 7 we
study the coherence legth and the space-time form of the
nucleons/nuclei in the black disk regime. In section 8 we discuss
the experimental consequences of our results. \par Some properties
of the dipole approximation in QCD and DIS we use are discussed in
appendixes. In appendix A we remind the reader the basic ideas of
the derivation of the dipole model, in particular we give a
detailed derivation of the momentum representation of the total
crosssection. In appendix B we remind the reader the basic
formulae of the AJM model. In appendix C we recalculate the
characteristic transverse momenta for the black disk regime, using
CTEQ6 distribution functions and two-gluon nucleon formfactor
parametrization from ref. \cite{Rogers1,Rogers2}. In appendix D we
present the parametrizations of the structure functions that we
use in this paper. In appendix E we present some interpolation
formulae that describe the data on median momenta as function of x
and $Q^2$ as described in tables 1,2.

\section{The target rest frame description.}
\par In this section we shall use the dipole approximation
to derive the momentum representation for the total cross-sections
of the DIS for both transverse and longitudinal photons. Within
the LO approximation the QCD factorization theorem allows to
express the total cross section of
 the scattering of  the longitudinally
polarized photon with virtuality $Q^2 \gg \Lambda_{QCD}^2$ off a
hadron target as the convolution of the square of the virtual
photon wave function calculated in the dipole approximation and
the cross section of the dipole scattering off a hadron
\cite{BFGMS,FS,FRS,Muellereikonal}. In the target rest frame the
cross section for the scattering of longitudinally polarized
photon has the form  (see  appendix A in this paper for the
detailed derivation of this formula):
\begin{equation}
\sigma(\gamma_L^{*}+T\to X)={e^2\over 12\pi^2} \int d^2 k_t dz
\left<\psi_{\gamma_L^{*}}(k_t,z)\right|
 \sigma(s,k_t^2) \left|\psi_{\gamma_L^{*}}(k_t,z)\right>.
\label{crosssection1}
\end{equation}
Here $\sigma$ is the dipole cross section operator:
\begin{equation}
\sigma= F^2\cdot \pi^2 \alpha_s(4k_t^2)(-\vec \Delta_t ) \cdot
\tilde x G(\tilde{x}= (M^2+Q^2)/s, 4k_t^2),
 \label{cross section}
\end{equation}
 $\vec \Delta_t$ is the  two dimensional Laplace operator in the
space of the transverse momenta, and $M^2$
is the invariant mass squared of the dipole. In the coordinate
representation $\sigma$ is just a number function, and  not a
differential operator as in the momentum representation. \par In
the leading $\log(x_0/x)$ approximation a similar equation arises
where the cross section is expressed in terms of convolution of
impact factor and unintegrated gluon density.  In practice, both
equations should give close results. Integrating by parts over
$k_t$ it is easy to rewrite Eq. ~\ref{crosssection1} with the LO
accuracy in the form where the integrand is
 explicitly positive:
\begin{equation}
\sigma(\gamma_L^{*}+T\to X)={e^2\over 12\pi^2} \int \alpha_s
(4k_t^2)  d^2 k_t dz  \left<
\nabla{\psi_{\gamma_L^{*}}}(k_t,z)\right| f(s, z, k_t^2)
\left|\nabla \psi_{\gamma_L^{*}}(k_t,z)\right>,
\label{crosssection2}
\end{equation}
here
\begin{equation}
f=(4\pi^2/3)\alpha_s(4k_t^2)
\tilde xG(\tilde{x}, 4k_t^2).
\end{equation}
In the derivation we use  the boundary conditions that follow from
the fact that the photon wave function  decreases rapidly
 in the  $k_t^2\to \infty$ limit
 and that the
contribution of small $k_t$ is the higher twist effect (see
Appendix A for details).

\par
The cross section of the interaction of the longitudinal photon can be rewritten in the form of spectral
 representation  by  explicitly
differentiating the photon wave function :
 \beq
\sigma_L(x,Q^2)=6\pi\frac{\pi\alpha_{\rm e.m.}\sum
e^2_qF^2Q^2}{12} \int dM^2 \int dz \alpha_s (4M^2z(1-z))
\frac{M^2}{(M^2+Q^2)^4}\cdot g(\tilde{x},M^2). \label{r10} \eeq
 Eqs.\ref{crosssection2}, \ref{r10} follow from the $k_t$
factorization theorem for high energy processes. Here
$g(x,Q^2)=xG(x,Q^2)$.Thus Eq.\ref{r10} is the generalization of
the DGLAP approximation to the domain of high energies  where the
processes with large and increasing with energy $M^2$ play an
important role.

\par Note that in Eq. \ref{r10} the dependence on $z$ comes explicitly
only in the argument of the coupling constant. Practically, for
the dominant configurations  $z\sim 1/2$ ( and hence $4k_t^2
\approx M^2$). Then the dependence on z in Eq. \ref{r10} is absent
and one can consider Eq. \ref{r10} (with z=1/2) as a spectral
representation of a electroproduction amplitude for longitudinal
photons.
 The spectral representation of the electroproduction
amplitude over $M^2$  is a general property of  a quantum field
theory at large energies, where the coherence length significantly
exceeds the radius of  the target T  \cite{Gribov,Yennie}.  The
DGLAP approximation ensures additional general property: the
smaller  size  of the configuration in the wave function of
the projectile photon leads to the smaller interaction with the
target, but this interaction  increases with the energy more
rapidly.
 Such spectral representation is not valid for transverse photons
due to the AJM effects, since the configurations with smaller z
dominate (see below), and the integrand in the corresponding
equation contains the contribution of different configurations
(contrary to the preQCD spectral expansions).
\par
Eq. \ref{crosssection1} can be equivalently rewritten in terms of
integration in $k^2_t$ and $z$ as
 \beq
\sigma_L (x,Q^2)=6\pi\frac{\pi\alpha_{\rm e.m.}\sum
e^2_qF^2Q^2}{12} \int dk^2_t \alpha_s (4k^2_t)z^2(1-z)^2
\frac{k_t^2}{(k^2_t+Q^2z(1-z))^4}\cdot g(\tilde{x},4k^2_t).
\label{r10a} \eeq where $\tilde x$ is given by
$(k^2_t/((z(1-z))+Q^2)/s$. Here we take into account explicitly
the (rather weak) $z$-dependence of the integrand.
 \par
 In the NLO  approximation the structure of formulae should be the same except  the appearance of the
 additional $q\bar q g, ...$ components in
the wave function of photon due to the necessity to take into account  the QCD evolution of the photon
wave function \cite{FS}.
\par
The similar derivation can be made for the scattering of
 transverse photon in configurations of spatially small size.
 In this case the contribution of small $k_t$ region (Aligned Jet Model contribution) is  comparable
 to the pQCD one.  The main interest in this paper is in the
 region of high energies (HERA and beyond) i.e.  sufficiently small $\tilde{x}$, and small $Q^2$,
 where pQCD contribution
dominates because of the rapid increase
 of the gluon distribution with the decrease of $x$.
 The pQCD contribution into the total crosssection
 initiated by the transverse photon  has the form:
 \begin{eqnarray}
 \sigma_T&=& 6\pi\frac{\pi\alpha_{\rm e.m.}\sum
e^2_q F^2}{12} \nonumber\\[10pt]
&\times &\int^1_0dz \int
dM^2\alpha_s(4M^2z(1-z))\frac{z^2+(1-z)^2}{z(1-z)}
\frac{(M^4+Q^4)}{(M^2+Q^2)^4}\cdot g(\tilde{x},
4M^2z(1-z)).\nonumber\\[10pt]
\label{r11}
\end{eqnarray}
The invariant mass representation for the transverse photon does not give objective information on the
 parton transverse momentum distributions. The reason is of course the well known fact, that there are
 configurations with large invariant masses, but small transverse
momenta $k^2_t$. These configurations are the perturbative
analogues of the AJ configurations. It is useful to rewrite Eq.
\ref{r11} in terms of integration over $k^2_t$ and  $z$. The
reason is that the median invariant mass and transverse momentum
for transverse photons are not connected like in the  longitudinal
photon case, and we need two
 different equations to determine them.
Cross section initiated by the transverse photon has the form:
 \begin{eqnarray}
 \sigma_T&=& 6\pi\frac{\pi\alpha_{\rm e.m.}\sum
e^2_q F^2}{12} \nonumber\\[10pt]
&\times &\int^1_0dz \int dk_t^2\alpha_s(4k^2_t)(z^2+(1-z)^2)
\frac{(k_t^4+Q^4z^2(1-z)^2)}{(k^2+Q^2z(1-z))^4}\cdot g(\tilde{x},
4k^2_t).\nonumber\\[10pt]\label{r12}
\end{eqnarray}
We include a
 contribution of the
aligned jet configurations by
  imposing a cut off in transverse
 momenta. In particular,
in the numerical calculations using Eq. \ref{r12} we introduced a
cutoff in the space transverse momenta $k^2_t=M^2z(1-z)\ge u, u\sim
0.35$ GeV$^2$. The contribution of smaller $k^2_t$ in the total
crosssection was calculated using the AJM model.

\section{The double logarithmic approximation.}
\par
In this section we analyze the new properties of the pQCD regime
within the double log  approximation. The advantage of this
approximation is that it will allow us  to perform some of the
calculations analytically. Other calculations will be made
numerically but using the expressions that are known analytically.

\par
In the double logarithmic approximation the structure functions  are given by \cite{DDT}
\beq
xG(x,Q^2)=\int dj/(2\pi i)(x/x_0)^{j-1}(Q^2/Q^2_0)^{\gamma (j)},
\label{2.10}
\eeq
where  the anomalous dimension is
$$\gamma (j)=\frac{\alpha_s N_c}{\pi (j-1)}.$$
To simplify the calculation we assume, the initial condition for the  evolution with $Q^2$:
 \beq
G(x, Q_0^2)=\delta (x-1). \label{KS1} \eeq In the saddle point
approximation one finds \cite{DDT}: \beq
xG(x,Q^2)=\frac{\log(Q^2/Q^2_0)^{1/4}}{\log(x_0/x)^{3/4}}\exp{\sqrt{4\alpha_s(Q_0^2)(N_c/\pi)\log(Q^2/Q_0^2)\log(x_0/x)}}.
\label{2.11b}
 \eeq
Structure function of a hadron is given by the convolution of this
kernel with the nonperturbative  structure function of a hadron in
the normalization point $Q^2=Q_0^2$. Note that $G$ is increasing
with $Q^2$ .This is the pQCD contribution where virtualities of
exchanged gluons are large.

In the analysis of energy dependence of the parton momenta it is
legitimate to neglect  the pre-exponential factor, since the absolute
value of G as well as the pre-exponential factor weakly influence
the transverse scale, and its evolution with energy: \beq
xG(x,Q^2)=\exp{\sqrt{4\alpha_s(Q_0^2) N_c/\pi
\log(Q^2/Q_{0}^2)\log(x_0/x)}}. \label{2.11} \eeq We shall
consider here the double logarithmic approximation without the
running coupling constant.
 Taking into account the
running coupling constant leads to similar results. We shall not
present them here, since they are more cumbersome, and are
essentially reproduced when we shall tackle the full LO
calculations in the next section.
\subsection{Energy dependence of the quark
transverse momenta for fragmentation processes initiated by
longitudinal photon.}
\par
We shall find analytically the scale of the transverse momenta in
the limit  where $s\gg M^2\gg Q^2$ . For certainty we restrict
ourselves to the contribution of light quarks.
\par
At large $Q^2$ the crosssection for the scattering of the
longitudinal photon is dominated by the contribution of the
spatially small dipoles, so it is legitimate to neglect the quark
masses.  In this limit the cross section is proportional to \beq
\sigma_L \propto Q^2 \int dM^2 n(M^2,s,Q^2), \label{dens} \eeq
where the function $n(M^2,s,Q^2)$ is given by  Eqs.~\ref{r10},
\ref{r10a}:
\begin{eqnarray}
n(M^2,s,Q^2)&=& \alpha_s(M^2/4) \frac{M^2}{(M^2+Q^2)^4}
\nonumber\\[10pt]&\times&
\exp(\sqrt{4\alpha_s(Q_0^2)(N_c/\pi)\log(M^2/Q_0^2)\log((x_0s/(M^2+Q^2))},\nonumber\\[10pt]
\label{2.11a}
\end{eqnarray}
where $x_0=Q^2/s_0$. Here we keep only large terms depending on
$M^2$ (we ignore the $M^2$ independent normalization factor
irrelevant for the calculations below).

\par
Let us show that the maximum of $n(M^2, s, Q^2)$  increases with the energy.
 At very high energies $n$ is proportional  to
 \begin{eqnarray}
n&\sim&
\exp(\log\alpha_s(M^2/4)+\log(M^2/Q^2)-4\log((Q^2+M^2)/Q^2)
\nonumber\\[10pt]
&+&\sqrt{4\alpha_s(N_c/\pi)(\log(s/s_0)-\log((Q^2+M^2)/Q_0^2))
\log(M^2/Q_0^2)}).\nonumber\\[10pt] \label{hen1} \end{eqnarray}
 In the limit of fixed $Q^2$ but very large energies,
 $\log(s/s_0)\gg \log((Q^2+M^2)/Q^2_0))$.  Let us assume that for the maximum:  $M^2\gg Q^2$.
 The
  maximum of the expression \ref{hen1} under this  assumption can be found analytically.
   We shall see that this assumption is indeed self-consistent. Indeed, under the latter assumption
 we can differentiate  Eq.~\ref{hen1} over $M^2$ and find for the
 maximum:
 \beq M^2= M_0^2(s/s_0)^{\alpha_s(N_c/\pi)/9}. \label{infu} \eeq
Here $Q_0^2\sim Q^2$ and $s_0\sim Q^2$. We
will refer to this extremum value of $M^2$ as  $M^2_1$.
We see from Eq. \ref{infu} that the condition $M^2_1>>Q^2$ is self-consistent
at very high energies.
\par
At the  extremum
  $n\propto
(\alpha_s(M_{1}^2/4)/M_1^6 \exp
((N_c/\pi)(\alpha_s/3)\log(s/s_0))$. Therefore \beq
\frac{d\sigma_L}{dM^2}\vert_{M^2=M^2_1}\approx
\alpha_s(M_1^2/4)(Q^2/M_1^6)(\exp((N_c/\pi)(\alpha_s(Q_0^2)/3)\log(s/s_0))
\label{red4} \eeq
\par
However, the position of the maximum of the integrand is not
sufficient to characterize the relevant transverse scales as a
large range of $M^2$ is important in  the integrand. In
particular, calculation of second derivative shows that dispersion
over $M^2=M_1^2$ is large.The width of the distribution over
$\log(M^2/M_0^2)$ is $\sqrt{2/3\log(M_1^2/M_0^2)}$.

Hence we need to determine $M^2$ range which gives most of
the integrand support. For certainty, we define the  range of $M^2 \le M^2_t$ which provides a fixed,
 say, $50\%$ fraction of the total perturbative cross section.  Let us estimate how this scale increases
 with the energy in the double log approximation. First,  let us consider the total
  cross section. The upper limit $u$ of integration over $M^2$ is
determined by the kinematic condition through $t_{\rm min}$,
giving that the allowed invariant masses  $M^2\ll s$. We choose
upper limit of integration as \beq M^2\le M^2_{\ max}=0.2s,
\label{m1} \eeq from the cross section of diffraction although the
result of numerical calculations is insensitive to the upper bound
because essential $M^2$ are significantly smaller. In fact the
integral for the cross section converges  long before the upper
limit of integration \ref{m1} is reached (see the discussion
below).
\par
Let us first calculate the median scale semi analytically. Within the double logarithmic approximation, and
 assuming that the conditions
$\log(s/s_0)\gg \log((Q^2+M^2)/(Q^2+M_0^2)$, is still
valid for the relevant $M^2$, the integral for the
 cross section can be written as:
\begin{eqnarray} \sigma (u)&=&
(Q^2/Q_0^4)\int^{\log(u/M_0^2)}_0d\log(M^2/Q^2_0)
\alpha_s(M^2/4)\exp(-2\ln(M^2/Q_0^2)\nonumber\\[10pt]
&+&\sqrt{(4\alpha_sN_c/\pi)\log(M^2/Q^2_0)\log(s/s_0)}).\nonumber\\[10pt]
\label{est}
\end{eqnarray}
Here $u$ is the upper cut-off in the invariant
masses. Introducing the  new variable $t=\log(M^2/Q^2_0)$, we obtain:
 \beq
 \sigma(u)=(Q^2/M_0^4)\int_0^{\kappa (u)} dt
\alpha_s(tM_0^2/4)\exp(-2t+\sqrt{(4\alpha_sN_c/\pi
)\log(s/s_0)t}), \label{s1} \eeq where $\kappa (u)=\log(u/s_0)$.
The integral for the total cross section is given by the equation
similar to Eq. \ref{s1}, with the upper integration limit being
replaced by $\kappa(s)=\sqrt{\log(0.2s/s_0)}$. Note that the
essential scale of integration is determined by exponent, and is
very weakly influenced by the exact value of a upper cut-off.  The
integral \ref{s1} is actually the  error function \cite{AS}, which
can be easily evaluated numerically.  Requiring that it
 gives one half of the  cross  section we find
  \beq
M^2_t\sim Q_0^2(s/s_0)^{0.28\alpha_sN_c/\pi}. \label{1} \eeq
Evidently, for sufficiently large $s$ our initial assumption
$\log(M^2/Q_0^2)\gg \log(Q^2/Q_0^2)$ is fully self-consistent.
This is because the decrease of $n$ with
 $M^2$ due to $1/M^6$  terms in the integrand of
Eq. \ref{r10} is partially compensated by the rising exponential,
giving a relatively slow decrease of $n$ to the right of the
maximum of the integrand.
\par
Note that the rate of the increase of $M^2_t$ with $s$ is much  higher than for $M^2_1$ due to the slow
 decrease of the integrand with $M^2$. The cross section of jet production with $M^2$ at this interval also
 increases with the energy as
 \beq
 \frac{d\sigma}{dM^2}\vert_{M^2=M^2_{\rm T}}\sim (s/s_0)^{0.24\alpha_sN_c/\pi}.
 \label{red301}
 \eeq
\par
The analytical calculations supply the  pattern for the behavior
of the transverse momenta in double log approximation. In order to
understand the dependence of the median scale on both the energy
and $Q^2$ quantitatively in the double logarithmic approximation
we made numerical calculation of the characteristic transverse
momenta  using the DGLAP double log structure function. We find
that the increase rate of the transverse
 momenta indeed does not depend on the external
virtuality $Q^2$. Considering the wide interval of energies and
$s=10^4\div 10^{11}$ GeV$^2$, and $20<Q^2<200$ GeV$^2$ we obtain
the approximate formulae: \beq M^2_t (x)\sim
0.7Q^2)\exp(0.17((4\alpha_sN_c/\pi )\log(x_0/x))^{0.55}).
\label{transverse11} \eeq Here $x_0=0.01$. We present this
estimate only for illustrative purposes, since the double
logarithmic approximation is  semirealistic only. Still our
results indicate that for external virtualities $Q^2<100 $ GeV$^2$
and energies which can be reached at LHeC the onset of a new pQCD
regime may take place.

The cross section of the jet production at this scale also
increases with the energy as
 \beq \frac{d\sigma}{dM^2}\vert_{M^2=M^2_{\rm t}}\sim (G(x,Q^2)/Q^6)(M^2_t(x)))\label{red3}\eeq
where $M^2_t(x)$ is given by Eq. \ref{transverse11}. The median
transverse momenta $k^2_t\sim M^2_t/4$, due to the dominance of
symmetric configurations.

\subsection{Transverse photon: the characteristic transverse scale in the photon fragmentation region.}

\par
The main difference between the longitudinal and transverse
structure functions in the DIS is the presence of the strongly
asymmetrical in $z$ configurations due to the presence of the
$(z(1-z))^{-1}$ factor in the spectral density. As a result there
is a competition between two effects.  One is  a slower decrease
of the spectral function with $M^2$ (by the factor $M^2/Q^2$),
leading to the  more rapid increase of the characteristic
transverse momenta for the symmetric configurations. Another
effect is the presence of the asymmetric ($z\to 0$) configurations
which are characterized by the small transverse momenta $k^2_t$
for a given invariant mass $M^2$. For such configurations the rate
of increase of the gluon structure function with energy is small.
\par Let us first show that the transverse momenta increase rapidly
for symmetric configurations.
The spectral representation for the transverse photon for
symmetric configurations is proportional to \begin{eqnarray}
n(M^2,Q^2,s)&\sim& \frac{M^4+Q^4}{(M^2+Q^2)^4}\nonumber\\[10pt]
&\times&\sqrt{4\alpha_s(N_c/\pi)(\log(s/s_0)-\log((Q^2+M^2)/(Q_0^2))
\log(M^2/Q_0^2)}).\nonumber\\[10pt] \label{t1} \end{eqnarray}
In the high energy limit, when $M^2_1\gg Q^2$, we find for the
dependence of the maximum of $n$ on energy: \beq M^2_1\sim
Q_0^2(s/s_0))^{\alpha_s(N_c/\pi)/4}. \label{t2} \eeq This is twice
as fast  increase as for the case of the longitudinal photon. $M^2_1$
increases with $s$ at high energies and thus the condition $M^2\gg
Q^2$ is perfectly self-consistent at very high energies.
In addition we can calculate the total cross section in the same
approximation semi-analytically getting the error function and
obtain the rate of increase $(s/s_0))^{0.14(4\alpha_sN_c/\pi)}$,
which is twice that for the longitudinal case.
\par
These two results are applicable to the symmetric configurations
only.
 On the other hand, at least at achievable  energies, the dominant contributions for transverse photon
crosssection are asymmetric, with z close to 0 or 1. In order to
take these configurations into account we performed a numerical
calculation  using
 the gluon  distribution function within the double log approximation. The  result is that the characteristic
  median scale $M^2_t$
increases as \beq M^2_t\sim
M^2(Q^2)(s/s_0)^{0.1(4\alpha_sN_c/\pi)}.
\label{transverse21}
\eeq
The value of the exponent is $0.12$ for the beginning
of the studied energy range  $s\sim 10^4 \div 10^{11}$ GeV$^2$, and   decreases to 0.09
 at the upper end (for typical $\alpha_s=0.25$).
Thus the rate of the increase with the energy is approximately the
same as for longitudinal photons for not very high energies. For
very high energies the symmetric configurations win over
asymmetric ones, leading to the increase of the average transverse
momentum squared which is   twice as large as in the longitudinal
case.
The precise determination of the scale $M_0^2(Q^2)$ is beyond the
accuracy of this approximation. Effectively we obtain the dependence
$M^2_t\sim 0.7Q^2 (x_0/x)^{0.1(4\alpha_sN_c/\pi)}$.
\par
One can also estimate the rate of the increase of the jet
production cross section : \beq d\sigma_T/dM^2_{\rm M^2_t}\sim
(xG(x,Q^2)/Q^4)(1/3)(1+0.5(x_0/x)^{0.24}). \label{df23} \eeq We
found  a rapid increase of the jets multiplicity with energy. Thus
the rate of the increase with energy of the transverse momenta of
quarks in the current fragmentation region for transversely
polarized photon is significantly more rapid. Consequently we find
that $\sigma_L/\sigma_T \approx \alpha_s Q^2/M^2$ being
numerically small should slowly decrease with energy at
sufficiently high energies .
\par
We conclude that it is possible to show analytically that for
very high (asymptotic) energies the relevant invariant masses
extend well  beyond $Q^2$ and increase with the energy.
The direct numerical calculation of the $M^2_t$ scale shows that the rate of increase
is independent of external virtuality.
\par Note that due to the significant contribution of the
asymmetric configurations the median transverse momenta is much
smaller than $M^2_t/4$. The simple numerical calculations using
$k^2_t=M^2z(1-z)$, shows that the average transverse momenta
(including nonsymmetric configurations) rapidly increases like
$a(Q^2)/(x/0.01)^{0.12}$, with the exponent once again is
independent of external virtuality, and $a(Q^2)\sim 0.6{\rm
GeV}^2+0.02Q^2$, i.e. $k^2_t$ is much smaller that $M^2/4$,
especially for small virtualities.

\section{The characteristic transverse momenta in hard
fragmentation processes in LO approximation.}
\par
In the previous section we discussed the quantitative dependence
of the characteristic transverse scales of DIS in the double
logarithmic approximation in the dipole model at unrealistically
high energies, where the analytical calculations are possible.
Here we carry out the calculations for realistic energies and
realistic structure functions. The numerical results indicate that
the effects discussed above are manifest even at the energies of
the order $s\sim 10^5\div 10^6$ GeV$^2$.
\par We will also consider the extrapolation of  our results to energies of the
order $s\sim 10^7 $GeV$^2$. These energies are unattainable  at
existing facilities. The proposed LHeC collider may reach the
invariant energies  of order $10^6$ GeV$^2$ \cite{Newman}.
 However these results are interesting  from the theoretical point of
view- probing the limits of the
pQCD.
\par
The challenging and unresolved problem  is how to use resummation
methods at extremely small $x$ \cite{ABF,Ciafaloni}
 to evaluate dependence on energy of parton distribution in the current fragmentation region.
   One can substantiate this point by  evaluation of the number of radiated gluons in the multiRegge
kinematics \cite{FSW}, obtaining 1-2 emissions at HERA energies. However  at extremely small x where number
 of gluon  radiations would be sufficiently large and therefore these models
can become  applicable. This interesting problem is beyond the scope of this paper.

\subsection{The longitudinal photons.}
\par
\begin{figure}[htbp]
\centerline{\epsfig{figure=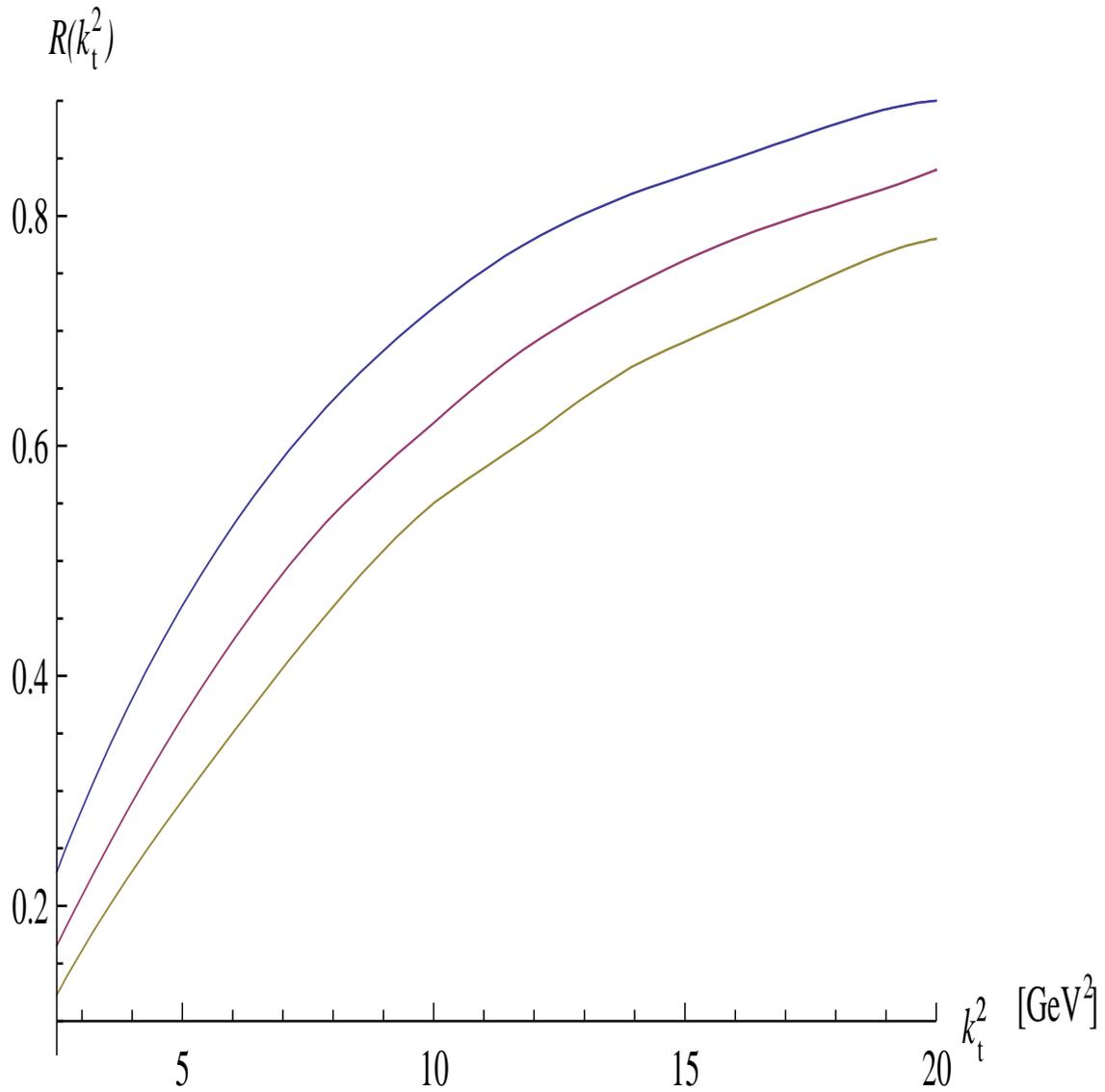,width=15cm,height=15cm,clip=}}
\caption{ The ratio $R(k^2_t)$  for $Q^2=40$ GeV$^2$ for
longitudinal photons. The three curves correspond to x=$10^{-3}$
(upper one), $10^{-5}$ (middle one) and $10^{-7}$ (lower one).}
\label{S2}
\end{figure}

In the case of longitudinal photons we have considered the
characteristic median/average transverse momenta scale, that
corresponds to the half of the total crosssection $\sigma_L$. This
scale is determined from Eq. \ref{r10} by first integrating over z
for given $k_t$, and then analyzing the corresponding jet
distribution.  The results of the numerical analysis using the
CTEQ6 gluon distribution functions are given in the Table 1. In
Figure\ref{S2}  we present the characteristic graphs for the ratio
\beq R(k^2_t)=\frac{\sigma( k^2_t)}{\sigma_L},\label{1000}\eeq
where $\sigma (k^2_t)$ corresponds to the result of integration of
Eq.\ref{r10a} over transverse momenta
 $\le k^2_t$. We see from Fig. \ref{S2},  that  for fixed $k_t$
 $R(k_t)$ slowly increases with the increase of the energy.
 The results based on using CTEQ5 parametrization
are qualitatively similar, although the increase of median $k^2_t$
with the energy is more rapid. The energy dependence
 of median $k^2_t$ can be described
with a very good accuracy  by an approximate formula
$(x/0.01)^{0.04+0.025\log(Q^2/Q_0^2)}.$
 Here $Q^2_0=10$ GeV$^2$,$x_0\sim 0.01$. The power increases from
$\sim 0.04$ at $Q^2\sim 5$ GeV$^2$, to $0.09$ at $Q^2\sim 100$
GeV$^2$. The interpolation expression which provides a good
description of the results presented in Table 1 is given in
Appendix E.
  For CTEQ5 the
exponent in Eq. \ref{tr1} increases to 0.1 at $Q^2=100 GeV^2$
instead of $0.09$. This is consistent with the enhanced rate of
the increase of CTEQ5 structure functions
as compared to the CTEQ6 ones (see below).
\par
These results allow us to estimate the scales, where one  expects
the appearance of the new QCD regime, i.e. one has to use the
$k_t$ factorization approach. Indeed, the DGLAP approximation is
based on the strong ordering in all rungs of the ladder, in
particular in the first rung (the impact factor in the $4k_t$
factorization language ) we must have $4\Lambda_{\rm QCD}^2\le
4k^2_t\le Q^2$. It is clear, this ordering can not hold,  once the
median $4k^2_{t}$ becomes of order $Q^2$. Then we obtain the
condition (using CTEQ6 distribution functions): \beq
4a(Q^2)/(x/0.01)^{0.04+0.025\log(Q^2/Q_0^2)}\sim Q^2.
\label{condition1} \eeq Here the function $a$  corresponds to the
transverse momenta at $x=0.01$.
\par
The analysis of Table 1 and Fig.\ref{S2} shows
that for $Q^2=5$ GeV$^2$ one gets from eq. \ref{condition1} $x\sim
10^{-4}$, for $Q^2=10$ GeV$^2$ one gets  $x\sim 10^{-6}$, which
may be reached at LHeC. For larger $Q^2$ we are however beyond the
realistic energies: say for $Q^2\sim 20$ GeV$^2$ we need $x\sim
10^{-9}$. The use of CTEQ5 gives qualitatively the same results
(for $Q^2=30 $ GeV$^2$ we obtain $x\sim 10^{-8}$.  Thus we may
hope to observe the onset
 of the new regime for  the $k_t$ dependence analyzing
  small $x$ jet distributions at LHeC/LHC.
\par
The similar analysis can be made using Eq. \ref{r10} and
considering the invariant masses $M^2$. We obtain that our results
for median invariant masses  are $M^2\sim (4\div 5)k^2_t$,
confirming the dominance of the symmetric configurations.

\subsection{Transverse photons.}
\par
\begin{figure}[htbp]
\centerline{\epsfig{figure=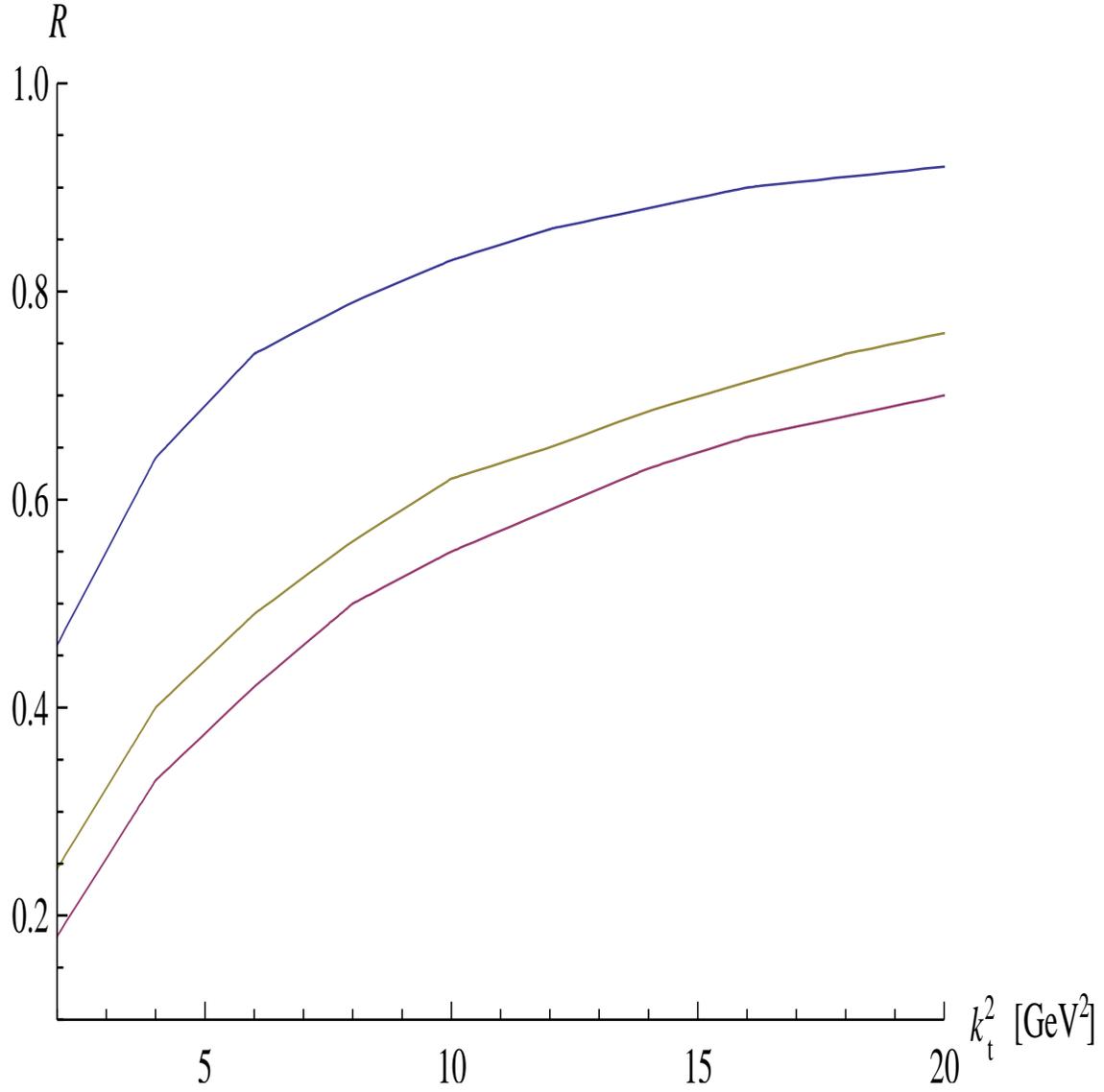,width=15cm,height=15cm,clip=}}
\caption{ The ratio $R(k^2_t)$  for $Q^2=40$ GeV$^2$ for
transverse photons. The three curves correspond to x=$10^{-3}$
(upper one), $10^{-5}$ (middle one) and $10^{-7}$ (lower one).}
\label{S5}
\end{figure}

We perform the numerical analysis for the transverse photons using
eqs. \ref{r11},\ref{r12} in the same fashion as for the
longitudinal photons. The results for the median transverse
momenta are given in Table 2. In figure \ref{S5}  we depicted the
characteristic function $R(k^2_t)$ given by Eq. \ref{1000} that
gives the characteristic momenta as a function of x for typical
value of $Q^2=40$ GeV$^2$. The characteristic energy dependence in
Table 2 for median $k^2_t$ is
$(x/0.01)^{0.09+0.014\log(Q^2/Q_0^2)}$ where $x_0= 0.01, Q^2_0=10$
GeV$^2$. A complete expression is given in appendix E. The curves
in Fig. \ref{S5} clearly show that the characteristic momenta
increase with the increase of $1/x$, as the corresponding curves
slowly shift to the right.
\par We see from the Table 2 that the average transverse
momenta for longitudinal photons is significantly larger than for
transverse photons. On the other hand, the invariant masses for
transverse photons are always significantly larger than $4k^2_t$.
This is due to the large contribution of the AJM type
configurations with $z\sim 0,1$ . Since $M^2=k^2_t/(z(1-z))$,   a
more slow increase  of $M^2$ than of $k^2_t$ is consistent with
the slow increase of average $z$ towards 1/2, i.e. the symmetric
configurations become dominant, but only at asymptotically large
energies.
\par Once again, we can estimate the boundary of the region
where the direct DGLAP approach stops being self-consistent.
Assuming $k^2_t\sim Q^2/4$, we obtain that the boundary for $Q^2 =
3, 5 , 10\,\, \mbox{GeV}^2$ is reached at $x \sim 10^{-3},
10^{-4}, 10^{-6}$. For higher $Q^2$ this boundary lies at
unrealistically high energies. The use of the CTEQ5
parametrization gives qualitatively the same results.
\par
So far we considered only perturbative QCD contribution, and the
median transverse momentum in Table 2 is determined relative to
the total perturbative cross section, i.e. the one starting from
the cut off $u=0.35$ GeV$^2$. However, as we mentioned above,
there is also important AJ  contribution in a total cross section.
 In this paper
we take them into account using the AJM model \cite{FS2} (see also Appendix B).
  This contribution is given in Table 3.
 It is well known that at low $x$ AJM gives
a dominant contribution into cross section. For $x\sim 0.01$, $Q^2\sim 1$ GeV$^2$
we see from Table 3 that AJM gives $70\%$ of the total cross section.
 However even at HERA
energies the contribution of AJM into the total cross section
remains significant.  Note that the median $k^2_t$ at small
virtualities at HERA energies significantly decreases if we
calculate it using the cross section that includes both the pQCD
and soft (AJM) contributions. For example, at $Q^2\sim 10$ GeV$^2$
the median transverse momentum squared  decreases by almost a
factor of two down to
 $k^2_t\sim 0.65$ GeV$^2$.
\section{Deeply virtual Compton scattering.}

\par
As the application of the formulae obtained in this paper we shall
consider the DVCS processes $\gamma+p\to \gamma^*+p$. We shall
show that the slow increase  in the median transverse momenta
leads to the slow decrease of the ratio $R=A_{\rm DVCS}/A_{\rm
CS}$ with energy to the limiting value  equal one.
\par
The DCVS amplitude is described in pQCD by the same formula
\ref{r11}  as the amplitude describing total cross section of DIS
at given  $x,Q^2$ but with the substitution in
Eq.\ref{crosssection2} of the wave function of virtual photon by
wave function of a real photon,  i.e. $Q^2=0$.

As a  result in  pQCD $R$ has the form : \beq R_{\rm
pQCD}=\displaystyle{ \frac{\int^1_0 dz \int dM^2 \alpha_s
(M^2z(1-z)) (1/(M^2+Q^2)^2)\cdot g(\tilde{x},M^2).}{\int^1_0 dz
\int dM^2 \alpha_s (M^2z(1-z)) ((M^4+Q^4)/(M^2+Q^2)^4))\cdot
g(\tilde{x},M^2).}} \label{DCVS} \eeq
\par
Let us note that strictly speaking, we must use the generalized
parton distributions  (GPD) in Eq. \ref{DCVS}.
 However the difference  between gluon GPD and gluon
pdf is not large in this case because fractions carried by gluons
in GPD differ by the factor $\approx$  two  at moderate x and tend
to one at extremely large energies as the consequence of increase
of parton momenta with energy. (In fact most of the  non-diagonal
effect in this approach is included  in the wave functions  of the
initial
 and final photons.)
Indeed, the  DVCS amplitude is expressed in terms of gluon GPD
where $x_{1}=M^2/s, x_{2}=(M^2+Q^2)/s$.
Our analysis above shows
 that $M^2\approx Q^2$ at moderate energies. So $x_1\approx
x\approx x_{2}/2$ . Analysis of the QCD evolution equation for GPD
\cite{GPD} shows that $x_1G(Q^2,x_1,x_2)=g(Q^2,x_1,x_2)\approx
g(Q^2,(x_1+x_2)/2)= g(Q^2, 1.5 x)$. At small $x$ $
g(Q^2,3/2x)/g(Q^2,x)\approx 1/(3/2)^{1/4}\approx 0.9$. In the
regime when $M^2\gg Q^2$ $x_1\approx x_2$. So GPD coincides with
the  gluon density. As a result we may neglect the difference
between GPD and distribution functions in the
considered kinematics.
 The numerical analysis of Eq. \ref{DCVS} shows
 that the ratio $R$ very slowly decreases with the increase of energy
due to  a slow increase of a ratio $M^2/Q^2$ discussed in the
previous section, and  $R\sim 1.6$ for HERA energies.
\par
The result Eq. \ref{DCVS} is however not complete since we
neglected the contribution of the AJ configurations.
Indeed, we already saw in the previos section (Table 3) that it gives
significant contribution in a total cross-section.
In the framework of the AJM model the ratio of amplitudes of the
DVCS to DIS can be calculated within the leading twist
approximation as \cite{FFS}: \beq R_{\rm AJM}=
\frac{Q^2+m_0^2}{Q^2}\log(1+\frac{Q^2}{m^2_0}).\label{DVCS1}\eeq
Here the parameter $m^2_0=0.3-0.5$ GeV$^2$ is the cut off
parameter $m^2_0\le m^2_\rho$, $m_\rho$ is the $\rho$ meson mass.
\par
Combining the pQCD and AJ model
 contributions we have
 \beq
 R=\frac{R_{\rm pQCD}\sigma_T+R_{\rm AJM}\sigma_{\rm
 AJM}}{\sigma_T+\sigma_{\rm AJM}}.
 \label{DVCS2a}
 \eeq
 \par
\begin{figure}[htbp]
\centerline{\epsfig{figure=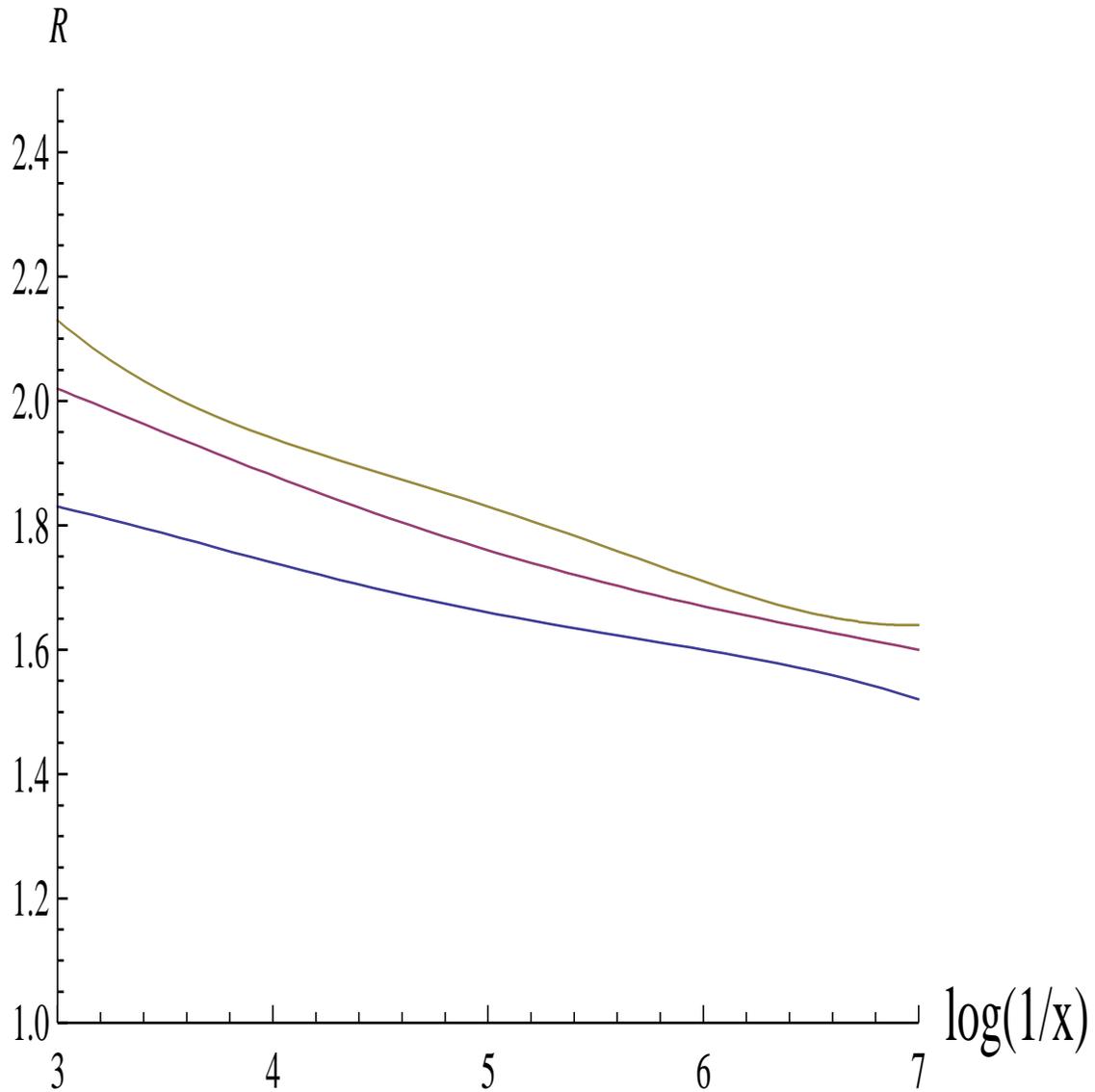,width=15cm,height=15cm,clip=}}
\caption{ The ratio R of the DVCS cross section to total
transverse cross section for different values of $Q^2$ as a
function of $x_B$. for $Q^2=5$ GeV$^2$ (lower curve), $20$ GeV$^2$
(middle curve), 60 GeV$^2$ (upper curve),the x axis corresponds to
$\log_{10}(1/x)$} \label{S8}
\end{figure}

 Here the pQCD contribution into the total cross section
 $\sigma_T$ is  given by Eq. \ref{r11} and the contribution of AJ to the  total cross section is
  given by AJM - Eq. \ref{AJM}. The results of numerical calculation with parameter $m^2_0\sim 0.5$ GeV$^2$ are given
in Table 4. We depict these results as a function of $x$ for
several values of $Q^2$ in Fig \ref{S8}.
 The ratio R is close to 2 at HERA energies and
increases with $Q^2$ (from 5 to 100 GeV$^2$ by $\sim 40\%$). This
result is in a good agreement with the analysis of the H1 and ZEUS
data in Ref. \cite{S}  (see in particular Table 4 in Ref.
\cite{S}). Our main prediction is that the ratio R should
decreases with the rise of energy. It tends to one at
asymptotically large energies
 in agreement with the result for the BDR \cite{PRL}.  However the onset of this regime is very slow.
This prediction can be checked experimentally in the study of DVCS
processes at LHeC.
\par
Our conclusion on the important role of AJM contribution in DVCS
at HERA energies is  in the qualitative agreement with the recent
experimental data \cite{LatestHera} that shows the important role
of soft QCD in the diffractive processes in DIS at HERA.
\par
We want to draw attention that agreement between experimental
results and theoretical prediction is rather good. This is due to
the fact that the interaction of dipole effectively includes the
NLO corrections since parton distributions were obtained by
fitting the experimental data. Consequently one may hope that NLO
corrections to impact factors are relatively small.
\par
Let us stress that the current calculation is preliminary.  More
detailed calculation should account for the contribution of
c-quark, and study in detail the dependence of R on the AJM
parameters).

\section{On the increase with energy  of parton momenta in the current fragmentation region in the black disc limit.}

\par
In the previous sections we found  the increase with the energy of
the parton momenta in the current
 fragmentation region in the pQCD regime.. Here we shall consider the configurations in the projectile wave function
that become dominant just near the transition to the black disk
limit. We
 will
explain that due to the dominance of these configurations, the
increase with the energy of the parton transverse momenta in the
current fragmentation region is further accelerated at the
energies  where the  DGLAP and BFKL approximations to  QCD
violate
 conservation  of probability/unitarity
\par The aim of the present chapter is to  analyze the properties of
this new QCD regime (we also call it the black disk regime; let us
stress that this regime is different from the possible new pQCD
regime that was discussed in the previous sections).  We shall
study the increase of the characteristic transverse momenta of the
configurations in the wave function of the projectile responsible
for the blackening, and derive a expression for the total
cross-section. This expression gives a total cross section as a
function of a transverse momentum increase rate.
 \par We will use
 the probability conservation for the amplitude describing the
dipole scattering at given impact parameter $b$. In the kinematics
of $x\ll 1$ Feynman amplitude
 describing scattering of dipole off a target $A(x,t, Q^2)$  is
calculable in terms of  partial amplitude $f(b^2,x,Q^2)$ as
\begin{eqnarray}
A(x,t=-q_t^2,Q^2)=2 s\int d^2b \exp(i\vec q_t\vec b)~ f(b^2,x,Q^2),
\label{Amplitude}
\end{eqnarray}
and
\begin{eqnarray}
f(b^2,x,Q^2)=\int {d^2q_t \over 8\pi^2} \exp(i\vec q_t\vec b){A(x,t,Q^2) \over s}.
\label{G}
\end{eqnarray}

\par
In a quantum theory the elastic cross section  for the scattering
of dipole off target T is given by the well known  formulae :
\begin{eqnarray}
\sigma_{el}=\int d^2b |f(b,x,Q^2)|^2,
\label{elastic}
\end{eqnarray}
The inelastic cross section is given by:
\begin{eqnarray}
\sigma_{in}=\int d^2b (1-\left |1+if(b^2,x,Q^2)\right|^2),
\label{inelastic}
\end{eqnarray}
and total cross sections is equal to:
\begin{eqnarray}
\sigma_{tot}=2\int d^2b  {\rm Im}~f(b^2,x,Q^2). \label{tot}
\end{eqnarray}

It follows from these equations that
\begin{eqnarray}
{\rm Im} f(b^2,x,Q^2) \le 1.
\end{eqnarray}
The limit
\begin{eqnarray}
{\rm Im} f(b^2,x,Q^2) =1, \label{BDL}
\end{eqnarray}
is reached in the small $x$ regime when a small transverse size of
the quark-gluon dipole is absorbed by the target with the $100\%$
probability at given $b$.

In the numerical evaluations it is convenient to use profile
function ${\rm Im} f=\Gamma$ and  neglect  ${\rm Re}  f$ as
compared to $Im f$, which is justified at sufficiently high
energies.

These equations allow us to evaluate some important properties of
the  onset of the new QCD regime of strong interaction with small
coupling constant and the kinematical region of applicability of
perturbative QCD to hard processes.

\begin{itemize}
 \item According to Eq.\ref{G} the dependence of $\Gamma$  on impact
parameter $b$ is given by  the Fourie transform of the two gluon
form factor of a nucleon.  It  was  measured in the hard exclusive
vector meson production at HERA and at FNAL. The two gluon
formfactor can be parameterized as: $F(q^2)=1/(1+q^2/\mu^2)^2$,
with $\mu$ decreasing with the decrease of $x$
($\mu^2(10^{-2})\sim 1 \,\,{\rm GeV}^2, \mu^2(10^{-4})\sim 0.7\,\,
{\rm GeV}^2$), cf. discussion in Appendix C. Within the framework
of pQCD at large $k_t$  the dependence of $\mu$ on $x$ due to
Gribov diffusion in the impact parameter space should be  a small
effect. Variation of $\mu$  comes mainly from the diffusion at
$Q^2\sim Q_0^2$ which is moderate, and it is further suppressed by
the DGLAP evolution \cite{Frankfurt:2003td}.
 Thus \beq
\Gamma(b,x) =(x_o/x)^{\lambda(4k_t^2)} \mu bK_1(\mu b).
\label{parametrization} \eeq
  Here
  $K_1$ is the MacDonald function and $k_t$ is the parton transverse momentum within the dipole.
 Parameter
  $x_o$
 is function of $k_t^2$ calculable in pQCD. The dependence on energy has been evaluated
 within pQCD  or using  the experimental data. Experimentally  $\lambda (10 GeV^2)\approx 0.2$
 and increases with $k_t^{2}$.

 \item Knowledge of $x, b, k_t$ dependencies of $\Gamma$ and
Eq.\ref{BDL} allows us to evaluate
 the characteristic $k_t$ in the region,   where the onset of the black disc regime occurs.
This evaluation predicts significantly more rapid  increase
 with energy of $k_t$ in the vicinity of black disc limit than
the calculations performed in the preceding sections. This is the
signature of the blackening of the interaction, i.e.  of the
transition to the regime described by  Eq.\ref{BDL}. Since the
integration over $k_t$ is logarithmic, it is determined  by the
upper limit of the integration over $k_t^2$ in the probability
defined by the wave function  of the dipole.

The dependence of $k_t$ on collision energy  is determined as the
rate of increase of the black disk momenta squared with energy:
\beq
 k_{t b }^2\sim k_{t b 0}^2(s/s_0)^{\kappa},
 \label{kappa}
 \eeq
where $s_0\sim 10^4$ GeV$^2$  (corresponds to the HERA energy).
The dependence of $k_{t b}^2$ on $s$ is depicted in Tables 5,6.
Note that $\kappa$ does not depend on whether we take $\Gamma=1/2$
or $\Gamma=1$ as a blackness condition, and is the same for
fermionic and gluonic dipoles. Our results show that this
coefficient weakly depends on energy, with $\kappa =0.3\div 0.4$
at the LHC energies and beyond.

\item The slope of $t$ dependence of the cross sections of hard
exclusive processes  at t=0 - $B(0)$ \beq B(0)=\frac{\int d^2b b^2
[\Gamma(b^2,x,Q^2)]}{\int d^2b [\Gamma(b, x, Q^2)]}, \eeq should
increase with energy in the vicinity of BDR and  at  larger
energies.  This property follows from the increase with energy of
the  essential impact parameters. Using Eq.\ref{G} we obtain
equation for
 maximal $b=b_{M}$  where $\Gamma$ is equal to 1:
 \beq
1/\log(\mu b_M K_1(\mu b_M))=\kappa \log (x_0/x).
\label{maximalb}
 \eeq
 At sufficiently large energies where
 $\kappa \log(x_0/x)\ge 1$  the above equation has  the solution:
 \beq
\mu b_M=\kappa \log (x_0/x). \label{maximalba}
 \eeq
Evidently $b_M(x\to 0)$ would become practically independent of
$x_0$.
\par Note that  within the LT approximation $\mu$ and therefore the slope
$B$ are slowly increasing with energy  in some interval of $x\sim
0.01$  as the consequence of spontaneously broken chiral symmetry.
The long range pion cloud carries small fraction $\approx
m_{\pi}/m_p$  of proton momentum  and therefore scattering off
this cloud is suppressed at not too small $x$ because amplitude is
rapidly decreasing with decrease of collision energy \cite{FSW}.
\par
A slightly more sophisticated reasoning which uses the
completeness over the hadronic states shows that
Eqs.\ref{maximalb}, \ref{maximalba}
 are valid even if we account for the gluons and quark-antiquarks pairs in the wave function of the dipole.

\item Knowledge of $b_M$ and Eq.\ref{tot} allows us to  calculate the
total cross section of the scattering of the spatially small
dipole. We  have $\Gamma=1$ for $b\le b_M$, $\Gamma(b\ge b_M) =
bK_1(\mu(x) b)/b_MK_1(\mu(x) b_M)$. Then we have
\begin{eqnarray}
\sigma &=&2\pi b_{M}^2 +2
\int d^2b\Theta(b-b_M)~ \Gamma(b,x,Q^2) \nonumber\\[10pt]
&=& 2\pi (b_{M}^2 +2(b_M/\mu)K_2(b_M\mu)/K_1(b_M
\mu)).\nonumber\\[10pt]
\label{gf}
\end{eqnarray}
Here to calculate the  second integral we use an identity between
special functions:
$$ \frac{d}{dx}(x^2K_2(x))=-x^2K_1(x)$$
The second term in the second line of eq. \ref{gf} is the
contribution of peripheral collisions.
 Thus cross sections of hard processes integrated over impact parameters would increase with energy forever.
 \par
Note that using eq. \ref{gf} we can estimate, what part of the
total cross section comes from the black disk regime, and what
part from pQCD. Indeed, estimating $b_M$ as a scale at which the
black transverse momentum is of order $\sim 1$ GeV, we have
$b_M\sim 4.5
 $ GeV$^-1$ for typical $x\sim 10^{-6}$, $Q^2\sim 10$ GeV$^2$. We  use a typical value of $\mu\sim 0.7$ GeV for
  these energies.
  Then the contributions of black disk to total cross section is of  the order $66\%$. Let us stress that here we
  assume $b_M\ll 1/(2m_{\pi})$
 which is a typical scale for the phenomenon of spontaneously broken chiral symmetry in the soft  QCD physics.
 \end{itemize}

\section{The coherence  length and the shape of the fast nucleon and nuclei.}
\subsection{The coherence length.}
\par
 In the previous sections we determined the  energy dependence of the effective  transverse scale at
  high energies. This result allows us to  evaluate the coherence length.The coherence length $l_c$ corresponds
to the life-time of the dipole fluctuation at a given energy
in the rest frame of the target.  The original suggestion of the existence of the coherence length in the
 deep  inelastic scattering  was first made by Ioffe, Gribov and Pomeranchuk \cite{GIP,Gribov}. It was found
already in the sixties within the parton model approximation by Ioffe \cite{Ioffe} that  the coherence
length at moderate $x$   is  $l_c\sim 1/2m_Nx$  i.e. it linearly increases with energy. In the
pQCD as a result of QCD evolution coherence length increases with energy more slowly:
\begin{equation}
l_c=(1/2m_Nx)(s_0/s)^{\lambda}.\label{co1}
\end{equation}
Such a behavior of $l_c$ with energy has been found previously in
the numerical calculations of structure functions in the target
rest frame \cite{KS,BFcoh,BFS}.

\par In a similar way, once we have the onset of the black limit
regime, the coherence length corresponding to the dominant
configurations in the process decreases even more slowly:
\begin{equation}
l_c=(1/2m_Nx)(s_0/s)^{\kappa}.\label{co2}
\end{equation}
Here $\kappa$ is the rate of the increase of the black disk
transverse momenta, given by eq. \ref{kappa}.

\subsection{The shapes  of  the fast nucleons and  nucleus.}

\par The results from the previous subsection have  the important consequences for the
transverse  structure of the hadrons and nuclei.
\par
Let us consider the longitudinal distribution of the partons in a
fast hadron.
In the parton model the longitudinal spread of the gluonic cloud
is $L_z\sim 1/\mu$ for
 the wee partons (where  $\mu\sim \, 0.3 \div 0.4 \, {\rm GeV/c}$
is the  mass scale of soft QCD processes.
) and it is much larger than for harder partons, with $L_z \sim 1/xP_h$
 for partons carrying a  finite $x$  fraction of the hadron momentum \cite{Gribovspace-time}.
 The picture is changed qualitatively in the
 limit of very high energies when interactions
reach BD regime for $k_t \gg \mu$. In this case the smallest possible characteristic momenta in the frame where
 hadron is fast are of the order $k_t (BDR)$ which is a function of both initial energy
 and transverse coordinate, $b$ of the hadron.  Correspondingly, the longitudinal size is $\sim 1/ k_t (BDR) \ll 1/\mu$.
 Note here that we are discussing  longitudinal distribution for typical partons.
 There is always a tail to the momenta much smaller than typical all the way down to
  $k_t \sim \mu$ which corresponds to the partons with much larger longitudinal size
   (a pancake of soft gluons corresponding to the Gribov's picture).  However at large
    energies at the proximity of the unitarity limit the contribution of the
    gluons with $k_t<k_{\rm tb}$ is strongly suppressed \cite{Mueller02}.
In the BDR this tail is suppressed by a factor
$ k_t^2/k_t (BDR)^2$ in addition to the phase factor \cite{FGMS}.
   In the color glass condensate model the suppression is  exponential \cite{Dumitru:2002qt}.

\begin{figure}[ht]
\centering\includegraphics[height=10cm,width=14cm]{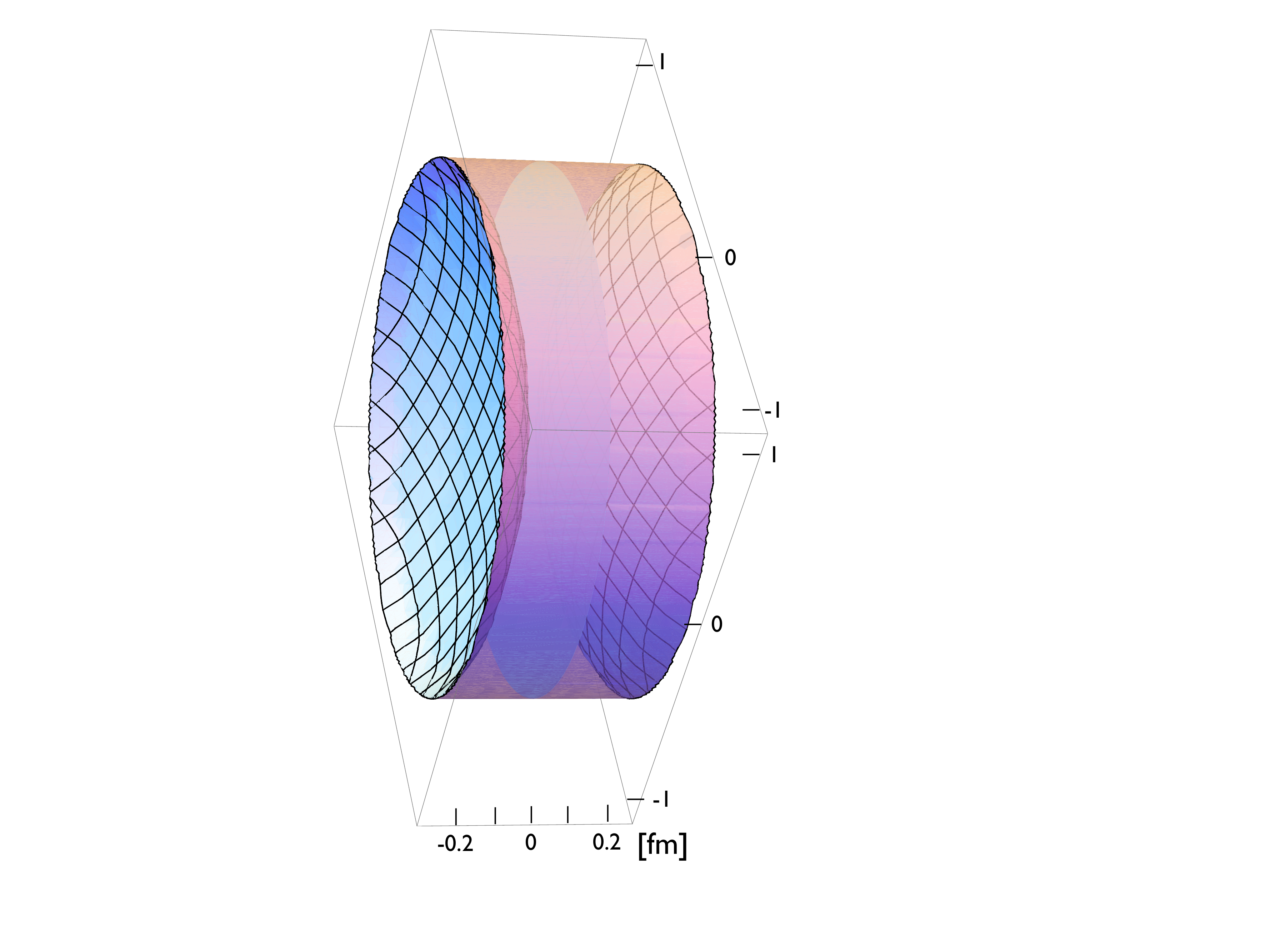}
\caption{3D image of the fast nucleon at $s=10^7  \, \mbox{GeV}^2$
and the resolution scale $ Q^2\le 40
 \mbox{GeV}^2$,} \label{S9}
\end{figure}

\begin{figure}[ht]
\centering\includegraphics[height=10cm,width=14cm]{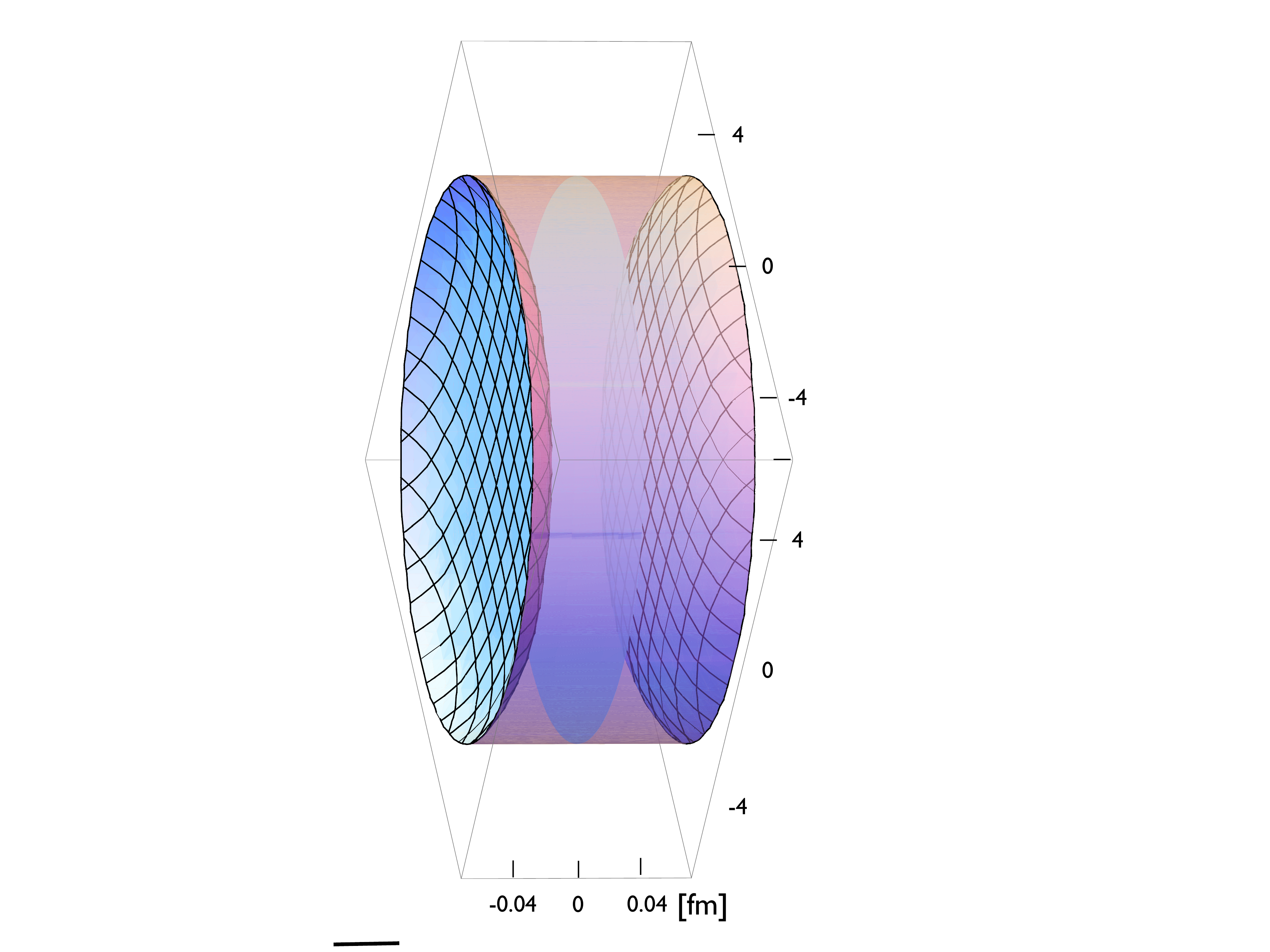}
\caption{3D image of the fast heavy nucleus (gold) at $s=10^7  \,
\mbox{GeV}^2 $ and the resolution scale $Q^2 \le  40
\mbox{GeV}^2$.Let us stress that the thickness of a nuclei in the Figure is much smaller
than the thickness of the nucleon.} \label{S10}
\end{figure}

Since the gluon parton density  decreases with the increase of $b$
the longitudinal size of the hadron is larger for large $b$, so a
hadron has a shape of biconcave lens, see Figs.
\ref{S9},\ref{S10}.

In the numerical calculation we took
\begin{equation}
\left| l_z \right|= 1/k_t(BDR),
\end{equation}
neglecting overall  factors of the order  of one (typically in the
Fourier transform one finds $\left<z\right> \sim {\pi \over
\left<p_z\right>}$).  We calculated $k_t(BDR)$ for fixed external
virtuality $Q^2 \sim 40 GeV^2$. Our results are not sensitive to
the value of $Q_0^2$, as the value of $Q^2$  only enters in the
combination  $x'= (Q^2+M^2)/s$, and the $k_t ^2$ we found were
comparable or larger than $Q^2/4$. Indeed, the direct calculation
shows that for small b the change of $1/k_t$ if we go between
external virtualities of $60$ and  $5$ GeV$^2$ is less than $5\%$.
Such weak dependence continues almost to the boundary of the
picture \ref{S9} where $k_t\sim 1$ GeV. Near the boundary the
uncertainty increase to $\sim 25\%$, meaning that for large b
(beyond those depicted in Fig. \ref{S9}) the nucleon once again
becomes a pancake and there is  a smooth transition between two
pictures ( biconcave lens and pancake). We want to emphasize here
that the discussed above weak dependence of  $k_t(BDR)$ on the
resolution scale indicates that the shape of the wave function for
small x is almost insensitive to the scale of the probe. 
\par
We depict the typical transverse quark structure of the fast
nucleon in Fig. \ref{S9}.  We see that it is drastically different
from the naive picture of a fast moving nucleon as a flat narrow
disk with small constant thickness. (Similar plot for the gluon
distribution is even more narrow). 
Note also that for the discussed  small x  range $k_t \ge 1 {\mbox
GeV/c}$ for $b\le 1 fm$. Since the spontaneous chiral  symmetry
breaking corresponds to quark virtuality $\mu^2 \le 1 GeV^2$,
probably $\sim 0.7 {\mbox GeV}^2$ \cite{DP}, corresponding to $k_t
\sim \sqrt{{2\over 3 } \mu^2} \sim 0.7 {\mbox GeV/c}$ the chiral
symmetry should be restored for a large range of $b$ in the proton
wave function for small x.
\par
Let us consider  the case of the DIS on the nuclei. First, we
consider the case of external virtualities of the order of several
GeV. In this case the  shadowing effects  to the large extent
cance the factor $A^{1/3}$ in the gluon density of a nucleus  for
the central impact parameters, $b$ \cite{FGS}, and
 the gluon density in the nuclei is comparable to that in a single nucleon for $b\sim 0$.
Consequently over the large range of the impact parameters the
nucleus longitudinal size is approximately the
 same as in the nucleon at $b\sim 0$.
 \par
 However for very small $x$
we find large $k_t(BDR)$ corresponding to $4 k_t^2(BDR) \ge  40$
GeV$^2$.  This is a self consistent value as indeed for such $Q^2$
the leading twist shadowing is small.
  Accordingly we calculated the shape of the nucleus for the external virtuality  $Q^2 \ge  40$ GeV$^2$.
   We should emphasize here that taking  a smaller virtuality would not significantly
 change our result for $k_t(BDR)$ (at the same time LT nuclear shadowing reduces a low momentum tail of
  the $k_t$ distribution
 as compared to the nucleon case).
   In the discussed limit of the small leading twist
 shadowing,  the corresponding gluon density
unintegrated over b is given by a product of a nucleon gluon density and the nuclear profile function:
\beq
T(b)=\int dz \rho(b,z),\label{bor1}
\eeq
where the nuclear three-dimensional
density is normalized to A. We use standard Fermi step
parametrization  \cite{Bohr}
 \beq
 \rho(r)=C(A)\frac{A}{1+\exp((r-R_A)/a)},\,\,\, R_A=
 1.1A^{1/3} {\rm fm}, a=0.56 {\rm fm}.\label{improved}\eeq Here  $r=\sqrt{z^2+b^2}$,
 and A is the atomic number.
C(A) is a  normalization factor, that can be
calculated numerically from the condition
$\int d^3r \rho(r)=A$.  At the zero
impact parameter $T(b) \approx 0.5 A^{1/3}$ for large A.
\par The dependence of the thickness of
a fast  nucleus as a function of the transverse size is depicted
in Fig.~\ref{S10} for a typical high energy $s=10^7$ GeV$^2$,
$Q^2=40$ GeV$^2$. We see that the nuclei also has a form of a
biconcave lens instead of a flat disk. The dependence on the
external virtuality for the nuclei is qualitatively very similar
to the case of the nucleon. For small b the dependence is very
weak (of order $5\%$) and increases only close to the boundary of
the biconcave lens region where it is of order $20\%$ ( and
$k_t\sim 1$ GeV). For larger b we smoothly return to the pancake
picture.
\par
Note that this picture is very counterintuitive: the thickness of
a nucleus is  smaller than of a nucleon in spite of
$\sim A^{1/3}$ nucleons at the same impact parameter.
The resolution of the paradox in the BD regime is quite simple:  the soft fields of
 individual nucleons destructively interfere  cancelling each other.  Besides for a
given impact
parameter $b $, the longitudinal size of a heavy nucleus
$1/ k_t^{(A)}(BDR) < 1/ k_t ^{(p)}(BDR)$ since the gluon distribution  function in
the nuclei $G_A(x,b) > G_N(x,b)$. So a naive classical  picture of a system build of the constituents being
 larger than  each of the constituents is grossly violated.
 The higher density of partons leads to the restoration of the chiral symmetry in a broad b range and much
 larger x range than in the nucleon case.

\section{Experimental consequences.}

\par
The current calculations of  the cross sections of the hard
processes at the LHC are based on the use of the DGLAP parton
distributions and the application of the factorization theorem.
Our results imply that the further analysis is needed to define
the kinematic regions where one can use DGLAP distributions. We
showed in the paper that for DIS at high energies
in the kinematical region of sufficiently small $x$ it is necessary
to use the $k_t$ factorization and the dipole model, instead of
the direct use of DGLAP.

A similar analysis must  be made for the $pp$ collisions at LHC.
It has been understood long ago that  the probability of pp
collisions at central impact parameter is close to 100\% (total
$\Gamma$ is close to 1) even for soft QCD, i.e. at lesser energies
than those necessary to achieve BDR for the hard interactions. The
compatibility of probability conservation with  the rapid increase
of hard interactions with energy , predicted by QCD, requires the
decrease of importance of soft QCD contribution with energy
\cite{FSZ}. As a result the hadronic state emerged in pp, pA, AA
collisions at sufficiently large energies consists of two phases.
Central collisions would be dominated by the strong interaction
with small coupling constant - the phase with unbroken chiral and
conformal symmetries. On the contrary, the  peripheral collisions
are dominated by the more familiar phase with broken chiral and
conformal symmetries.
 At these energies the QCD phase at central
 collisions - with the unbroken chiral and conformal symmetries -will be different from that
for the peripheral collisions.  This new phenomenona may appear
especially important for the central heavy
 ion collisions at LHC and at RHIC. Quantitative analysis
of this problem will be presented elsewhere.

\par
The hard processes initiated by the real photon can be directly
observed in the ultrapheripheral collisions \cite{Mark1}. The
 processes where a real photon scatters on a target,
 and creates two jets with an invariant mass $M^2$, can be analyzed in the dipole model by formally putting
 $Q^2=0$, while $M^2$ is an invariant mass of the jets. In this case  with a  good accuracy the spectral density
discussed above will give the spectrum of jets in the
fragmentation region. Our results show that
 the jet distribution over the transverse momenta will be broad
 with the maximum moving towards larger
transverse momenta with increase of the energy and centrality of
the $\gamma A$ collision.
\par We have seen that our results can also describe DCVS
processes. The ratio R of DCVS $\gamma^*\rightarrow \gamma^*$ and
forward amplitudes at $t=0$ is of order 2 at HERA energies at
small external virtualities, and rapidly growing with $Q^2$. This
ratio slowly decreases with the decrease of $x$.
\par Finally, our results can be checked directly, if and when the
LHeC facility will be built at CERN. One of us, B.Blok, thanks
S.Brodsky for the useful discussions of the results obtained in
the paper. This work was supported in part by the US DOE Contract
Number DE- FG02-93ER40771 and BSF.
\appendix
\section{The basics of the  dipole approximation approach.}
\par
The QCD factorization theorem allows
to express the total cross section of the DIS off a hadron through
the  convolution of the virtual photon wave function.
LO approximation is equivalent to
the dipole approximation and the cross section of the dipole
scattering off a hadron \cite{FS}.
\par
In this section we shall remind the reader the basic idea of the
derivation of this result \cite{FS,FRS}. Consider for the
simplicity the case of the longitudinal photon. The total cross
section of the photon scattering off the target $T$ has the
following factorized form:
 \beq \sigma =\int
\frac{d^4k}{(2\pi)^4}\frac{1}{s(k^2)^2}(2{\rm
Im}T^{abN}_{\mu\nu})({\rm Im}T^{ab}_{\bar\mu\bar\nu})
d_{\mu\bar\mu}d_{\nu\bar\nu}.\label{r1}\eeq Here
$d_{\mu\bar\mu}/k^2$ is the propagator of the exchanged gluon in
the light-cone gauge, $T\Gamma$ is the sum of the box diagrams
describing the scattering $\gamma^*g\rightarrow \bar qq$, and $T$
is the amplitude of the gluon scattering of the target.
Restricting ourselves to the contribution of longitudinal  gluon
\cite{Gtrick} polarization and using the Sudakoff decomposition,
the latter equation can be rewritten as \beq \sigma=
\int\frac{d\alpha d\beta d^2k_t}{8(pq)^2(2\pi)^4(k^2_t)^2}({\rm
Im} N^{ab}_{\mu\nu}p^{\mu}p^{\nu})({\rm
Im}T^{ab}_{\bar\mu\bar\nu}q^{\bar\mu}q^{\bar\nu}). \label{r2} \eeq
Using the Ward identity we can rewrite: \beq \frac{{\rm Im}
N^{ab}_{\mu\nu}p^{\mu}p^{\nu}}{4(pq)^2}=\frac{{\rm Im}
N^{ab}_{\mu_t\nu_t}k^{\mu_t}k^{\nu_t}}{(\beta s)^2}. \label{r3}
\eeq Since we integrate over $d^2k_t$, the r.h.s. of Eq. \ref{r3}
can be substituted by $\sum_{\mu_t=1,2}{\rm Im}
N^{ab}_{\mu_t\nu_t}k^2_t$ The latter term can be directly
connected with the scattering cross section  $\gamma^*g\rightarrow
\bar q q$: \beq
 \delta_{ab}(\beta s -Q^2)\sigma(\gamma^*g\rightarrow \bar qq)=\frac{1}{2}\sum_{\mu_t\nu_t}{\rm Im}N^{ab}_{\mu_t\nu_t}
\label{r4a} \eeq Here $\alpha,\beta $ are the Sudakoff parameters
of the gluon momenta k: \beq k=-\alpha q'+\beta
p'+k_t,\label{r4}\eeq q ia the photon and p is the target momenta.
The total cross section then can be rewritten as (see ref
\cite{FRS} for details)  \beq \int
\sigma_{\gamma^*g}\frac{d\beta}{\beta} \beta G_T(\beta,Q^2).
\label{r5}\eeq Thus the integrand in Eq. \ref{r2} for the total
cross section is positively defined since it is a product of the
cross section of the creation of $\bar q q$ pair and the structure
function of a target. Here $\beta=(Q^2+M^2)/s$.
\par
It is convenient to rewrite Eq. \ref{r5} by expressing  the cross
section $\sigma (\gamma^*g\rightarrow \bar q q)$ through the
light-cone wave functions of the virtual photon:
\begin{eqnarray}
 \sigma &\sim &\int \frac{d^2 k_t}{2(2\pi)^3}\int d^2r_t dz
 \frac{1}{2(2\pi )^3} \nonumber\\[10pt]
&\times&  \psi(r_t,z)(2\psi (r_t,z)-\psi (r_t+k_t)-\psi
(r_t-k_t))\nonumber\\[10pt] &\times&
\frac{4ImT^{ab}_{\mu_1\mu_2}k^{\mu_1}k^{\mu_2}}{s}.\nonumber\\[10pt]
\label{8.1}
\end{eqnarray}
Here $r_t$ is the transverse  momentum of the constituent within
the dipole and  $k_t$  is the transverse momentum of the exchanged
gluon. The tensor $T^{abT}$ is the sum of the diagrams describing
imaginary part of the amplitude for  the gluon scattering off the
target T. The integral of $T^{abT}$  over $d^{2} k_t$ is
proportional to the  gluon structure function of the target T. The
function $\psi$ is the wave functions of the virtual photon . Here
$r_t$ is transverse momentum
\par
Note that the validity of the dipole approximations is becoming
better for the larger transverse momenta of the dipole
constituents.
\par
Since $r_t\gg k_t$ we can expand the wave functions in the Taylor
series: \beq 2\psi (r_t,z)-\psi (r_t+k_t)-\psi (r_t-k_t))=-\psi
(r_t)\triangle \psi(r_t)k^2_t. \label{r6}\eeq The latter equation
is just the equation \ref{crosssection1}: \beq
 \sigma\sim \int dz\int d^2r_t\psi (r_t)\triangle\psi
 (r_t)g((M^2+Q^2)/s, M^2).
\label{r7} \eeq Here $r^2_t/(z(1-z))=M^2$ and $z$ is the fraction
of photon momentum carried by one of the constituents of the
dipole. \par The resulting cross section can be written either in
the momentum representation, giving the representation of the
total cross section as the dispersion integral over the invariant
masses of the dipoles, or in the coordinate representation, giving
the cross section as the integral over the transverse dipole
scales.
\par Similar derivation can be made for the scattering of the
spatially small transversely polarized photon.
Principal difference is that  the contribution of large
size q$\bar q$ configuration is not suppressed in the case of transversely polarized photon -leading to the
aligned jet model.
\par
Let us show now that within the leading logarithmic approximation one
can rewrite Eq. \ref{r7} in the explicitly positive form. Indeed,
 let us perform the integration over $d^2r_t$ by parts,
using $\triangle=\partial_i\partial_i, i=1,2$.  The resulting
expression contains four terms:
\begin{eqnarray}
 \sigma&\sim& \int dz (\int d^2 r_t
\partial_i\psi\partial_i\psi g((Q^2+M^2)/s, 4r_{t}^2)+\int d^2r_t\partial_i\psi\psi
(r_t)\partial_ig((M^2+Q^2)/s, 4r_{t}^2)\nonumber\\[10pt]
&+&{\rm boundary\,\,\,\,\, terms\,\,\,\,\,  from\,\,\,}
r_t\rightarrow 0+{\rm boundary\,\,\,\,\, terms,\,\,\, from}
\,\,\, r_t\rightarrow\infty .\nonumber\\[10pt]\label{r8}
\end{eqnarray}
Here for the convenience we use definition: $g=xG$. Let us compare
the relative contributions of the second and first terms. The
ratio of the corresponding integrands is determined by the ratio
of derivatives of the logarithms of the wave function and of the
structure function.
 The differentiation $(d/d(r^2_t)$ removes one
 $\log(r^2_t)$ from the pQCD series over the powers of
 $\alpha_s\log(r^2_t)$ describing the pQCD
 evolution of parton distributions. As a result the
logarithmic derivative of parton distribution is suppressed by
$\alpha_s$.
 Hence in the leading logarithmic approximation
the second term is parametrically smaller than the first one and
can be neglected .  Now consider the boundary term. The first of
the boundary terms corresponds to the contribution of the small
momenta (of $r_t\rightarrow 0$), i.e. to the aligned jet model
contribution. This contribution is known to be small for
$\sigma_L$. (It is zero within the parton model approximation).
The second boundary term corresponds to the contribution of the
very small dipoles, and clearly is zero since the
$\psi\partial_i\psi G\rightarrow 0$ as $r_t\rightarrow \infty$. We
conclude, that in the leading
 logarithmic approximation only the first term should be
 retained, and we come to Eq. \ref{crosssection2} in section 2 in
 the main text.
 \section{The aligned jet model (AJM)}
 The purpose of this appendix is to remind a reader the basic
 formulae
 used to calculate the
contribution of the AJM model to a total DIS cross sections and
forward amplitude of DVCS process in the section 4. The
contribution of the aligned jet model to the total cross section
initiated by the transversely polarized  virtual photon has been
evaluated in \cite{FS2}: \beq \sigma_T=\int_0^{0.7s}dM^2
\frac{M^2}{(M^2+Q^2)^2}\sigma_{ VN}(s/s_0)^{0.08}(\theta
(M_0^2-M^2)+\frac{3u}{M^2}\theta({M^2-M^2_0})). \label{AJM} \eeq
Here $u$ is the upper cut off, $M^2_0$ is a characteristic
hadronic scale of the order 1GeV$^2$ . The first term in the
formula corresponds to the contribution of the vector meson
dominance model and it is HT effect, while the second term is the
aligned jet model contribution. Since our interest in this
Appendix is focussed on moderately large $Q^2$ we use
$R(M^2)\equiv \rho(e^+e^-\rightarrow {\rm
hadrons})/\rho(e^+e^-\rightarrow \mu^+\mu^-))=2$. The contribution
of charm is not large but requires special treatment We assume
that the cross section for the interaction of hadronic size q$\bar
q$ pair with a nucleon target depends on collision energy as the
soft Pomeron exchange, that the AJM and meson - nucleon cross
sections are equal. The cross section of vector meson-proton
interaction -$\sigma_{Vp}$ is taken to be $\sim 25$ mb at the
energies of the order $s_0\sim 400$ GeV$^2$.
\par The contribution of the AJM model to the longitudinal model
cross sections is suppressed by a factor $4*u/Q^2$, i.e. \beq
\sigma_{\rm LAJM}\sim \frac{4u}{Q^2}\sigma_{\rm TAJM}\eeq
\section{The characteristic transverse momenta in the black disk
limit.}
 \par
 Let us now evaluate the scale of the black disc
limit where the partial wave for the forward scattering amplitude
reaches 1. This scale gives us the biconcave shape of the nucleon,
as it is explained in the section 5.
\par
The black disk scale was discussed already in a number of papers
(see e.g. Ref. \cite{Rogers2} for the recent discussion and
references ). Here we remind for a reader the rules of the
calculation of the black scale in the dipole model, discuss it's
dependence on energy. In order to calculate the corresponding
$M^2$ we need to calculate the partial wave for a dipole -target
scattering at zero impact parameter and then impose a black disc
limit condition: the partial wave should be 1. (The scale of black
disc limit depends on the impact parameter, so in the analysis
below we shall consider the case of the zero impact parameter, for
larger impact parameters the black disc limit arises at higher
energies (see section 6).
\par
Let us define in the standard way the profile function for the
dipole scattering: \beq \Gamma (s,b)=\frac{1}{2is(2\pi)^2}\int d^2
q_t\exp(i\vec q\vec b)A(s,t)\label{l1}\eeq Here $t=-q^2_t$, and
A(s,t) is the amplitude of the dipole scattering on the nucleon
with the momentum transfer q. The elastic, inelastic and total
crosssections are connected with this
 profile function as
\begin{eqnarray}
\sigma_{\rm tot}(s)&=&2\int d^2b {\rm Re}\Gamma
(s,b)\nonumber\\[10pt]
\sigma_{\rm el}(s)&=&\int d^2b \vert\Gamma
(s,b)\vert^2\nonumber\\[10pt]
\sigma_{\rm in}(s)&=&\int d^2b (2{\rm Re}\Gamma (s,b)-\vert\Gamma
(s,b)\vert^2).\nonumber\\[10pt]
\label{l2}\end{eqnarray}
\par The amplitude of dipole
nucleon scattering can be expressed through the total cross
section and gluon density \beq
A(s,t)=is\sigma(s)F_g(t).\label{l3}\eeq
 Here $\sigma (s)$ is given by Eq. \ref{1.3} with $F^2=4/3$ for
 the fermion dipole and $F^2=9/4$ for the gluonic dipole.
The two-gluon form factor of the nucleon
can be parametrized as to satisfy mainstream ideas on nucleon form factors:
 \beq
F_g(t)=\frac{1}{(1-t/m^2_g)^2}.\label{l4}\eeq
 \par
For the numerical dependence of the two-gluon nucleon form factor
parameter $m^2_g$ we shall use the parametrization \cite{FSW}
 \beq
m^2_g(x,M^2)=m^2_g(x,Q_0^2)(1+1.5\log(Q^2/Q_0^2))^{0.009\log(1/x)},\label{par1}\eeq
where
 \beq m^2_g(x,Q^2_0)=8/C,C={\rm max} (0.28 {\rm fm}^2,0.031
fm^2+0.0194 \log(0.1/x){\rm fm}^2),\label{par2}\eeq and $1
fm^2$=0.04  GeV$^{-2}$. Qualitatively, $m^2_g\sim 1$ GeV$^2$.
\par
This parametrization is based on the fit of cross sections of hard
diffractive processes in the kinematical domain: $x_B\sim
10^{-4}-10^{-1}$ and $Q^2$up to $10^4$ GeV$^2$ made in
ref.\cite{FSW} and was used in ref. \cite{Rogers2} for the
analysis of LHC collisions.
 \par
 Since absorption can not exceed 100\%  :
 \beq \Gamma (s,b)\le 1\label{l6}\eeq
 Let us make the Fourier transform
 of the form factor \ref{l4} into
 impact parameter space :
 \beq F(b)=(1/2\pi)^2 (\int d^2q\exp(i\vec q\vec
b)(m^4_g/(m_g^2+q_t^2)^2= (1/2\pi)^2\frac{m^3_gbK_1(m_gb)}{4\pi
}\label{1.5}\eeq Then
the condition of probability conservation (of 100\% absorption)
can be rewritten as
\beq \Gamma (s,b)=\frac{\sigma
(s)m^2_g}{4\pi}\frac{m_gb}{2}K_1(m_gb)\le 1\label{l9}\eeq
\par
Note that described above approach to the probability conservation differs
 from the saturation models which explore formulae of LO BFKL approximation
 derived for hard processes with one scale  and accounted for the conservation of probability
 in terms of  elastic eikonal approximation. ( For the recent review see \cite{CGC,Mueller}) .Our
 approach is based on the DGLAP formulae derived for
two scale (hard-soft scales) processes like total cross section of DIS and on the condition of
 complete absorption whose validity does not require applicability  of eikonal approximation.
Note also that in QCD inelastic diffraction in the scattering of spatially small dipole is
 significantly larger than elastic one \cite{FSW}.  For many phenomena (but not for all)  both  approaches
  lead to qualitatively
similar but quantitatively different predictions . Another advantage of the approach used in the paper
 is in the exploring impact parameter
gluon distribution which follows from the two gluon form factor
measured in the hard diffractive processes \cite{FSW}.

\par
The typical dependence of the BDR onset scale $k^2_b$  on energy
is presented in Table 5 for the  gluonic dipole scattering off a
proton target, and in table 6 for fermionic dipole, where we used
a CTEQ5 structure functions, the use of CTEQ6 leads to similar
results. We consider two possible criterions for reaching the
black disk limit: one when the partial wave $\Gamma $ at the
central impact parameter reaches 1, another when it reaches 1/2
\cite{FSW}. Indeed, when the partial wave reaches 1/2 the
probability of inelastic
 interactions reaches 3/4, i.e. interactions become strong and pQCD can not be used any more \cite{FSW}.

\par Let us note that
the interaction of quark dipoles (see table 6) is far from the
black disk regime for HERA energies, although nonperturbative
effects (corresponding to $\Gamma=1/2$) seem to start to appear
for $k_t<1 $ GeV.   The results in tables 5,6 are in good
agreement with the previous determination of these scales
\cite{Rogers1,Mark1,FSW}.
\par
For our purposes it will be important to determine
 the rate of increase of the $k^2_b$. This rate does not depend on the type of the dipole (fermionic or gluonic)
  or on the partial wave condition.
\par
For the gluonic dipole in DIS for s up to $s\sim 10^7$ GeV$^2$ the
BDR scales increase relatively slow as $s^{0.27}$ for CTEQ6 and
$s^{0.3}$ for CTEQ5 parametrisation. However at the energies above
$s\sim 10^7$ GeV$^2$ the increase speeds up with $M^2\sim s^n,
n={0.35}$ starting from $s\sim 10^8$ GeV$^2$ energies , n=0.32 for
the exponent for CTEQ6 parametrisation. The CTEQ5 parametrisation
leads to even more rapid increase with the energy with $n\sim
0.4$.

\section{The parametrisation of the gluon distribution function}

\par Throughout this paper we use the following approximate
formulae for CTEQ5L and CTEQ6L distribution functions. \par  The
CTEQ5L distributions has been shown to be in good agreement with
HERA data \cite{Rogers2}, in the range  of $x\sim
10^{-3},10^{-4}$. To extrapolate to very small x, we shall use the
approximate formulae in the form:
 \beq
g\sim a(M^2)/x^{c(M^2)}. \label{n1} \eeq Here the functions \beq
a(M^2)= 2.00123 - 1.69772\cdot 10/M^2 + 3.07651/\sqrt{M^2/10.} -
0.228087\cdot \log{(M^2/10)}, \label{n2a} \eeq \beq
c(M^2)=0.045\log(M^2)+0.17, \label{n3a} \eeq where $M^2$ is in
GeV$^2$. This formula is also the good fit to the observed
behavior of the structure functions
 measured
at HERA for $150 {\rm GeV}^2\ge
Q^2\ge 3$ GeV$^2$ , made by ZEUS and H1 collaborations \cite{H1}.
\par For CTEQ6 we have similar approximation, but with
\beq a(M^2)= 3.96 - 0.032\cdot 10/M^2 -1.82/\sqrt{M^2/10.} -
0.527\cdot \log{(M^2/10)}, \label{n2} \eeq
 \beq
c(M^2)=0.035\log(M^2)+0.2,
\label{n3}
\eeq
\par
The CTEQ6 distribution is qualitatively similar to CTEQ5
but grows slightly more slow, and is in better agreement with the
$J\Psi$ production data.
\section{Approximate formulae for the median transverse momenta.}
\par For longitudinal photons it is easy to see that the
results in a Table 1 with a very good accuracy can be described by
an approximate formula
\begin{eqnarray}
4k^2_t&=&\frac{a(Q^2)}{(x/0.01)^{0.04+0.02\log(Q^2/Q_0^2.}},\nonumber\\[10pt]
a(Q^2)&=&0.6{\rm GeV}^2+0.1 Q^2.\nonumber\\[10pt]
\label{tr1}\end{eqnarray} Here $Q^2_0=10$ GeV$^2$.
\par For the transverse photons the analysis of Table 2 gives an approximate formula
\begin{eqnarray}
k^2_t&=&\frac{a(Q^2)}{(x/0.01)^{0.09+0.014\log(Q^2/Q_0^2)}},\nonumber\\[10pt]
a(Q^2)&=&0.8 {\rm GeV}^2+0.025 Q^2.\nonumber\\[10pt]
\label{tr2a}\end{eqnarray} The invariant masses increase slightly
more slowly, with $M^2(x\sim 0.01)\sim 0.95Q^2+9$GeV$^2$, and
$M^2\sim 1/(x/0.01)^{0.07+0.015\log(Q^2/Q^2_0)}$, Here $Q_0^2=5$
GeV$^2$. If we use the CTEQ5 we obtain qualitatively the same
results as in Eq. \ref{tr2a} but the exponent is slightly larger
$4k^2_t\sim 1/x^{0.08+0.001\log(Q^2/Q^2_0)}$, and the invariant
masses increase also more rapidly, like $M^2\sim (0.95Q^2+9 {\rm
GeV}^2))/(x/0.01)^{0.08+0.015\log(Q^2/Q^2_0)}$.

\newpage
\begin{table*}[htbp]
\begin{center}
\begin{tabular}{|l|l|l|l|l|l|l|l|l|}\hline
 &$Q^2=3$ GeV$^2$&$5$ GeV$^2$&$10$ GeV$^2$&$20$ GeV$^2$&$40$
 GeV$^2$&$60$ GeV$^2$&$80$ GeV$^2$&$100$ GeV$^2$
 \\  \hline
 $s=10^{-3}$ GeV$^2$&0.82 GeV$^2$& 1.1 GeV$^2$&1.8 GeV$^2$ &3.14 GeV$^2$&5.5 GeV$^2$ &7.75 GeV$^2$&9.9 GeV$^2$&11.9 GeV$^2$\\ \hline
 $s=10^{-4}$ GeV$^2$&0.90 GeV$^2$& 1.25 GeV$^2$&2.1 GeV$^2$&3.65 GeV$^2$&6.5 GeV$^2$&9.3  GeV$^2$& 11.9 GeV$^2$& 14.5 GeV$^2$\\ \hline
 $s=10^{-5}$ GeV$^2$&0.95 GeV$^2$& 1.35 GeV$^2$&2.3 GeV$^2$&4 GeV$^2$&7.2 GeV$^2$& 10.2 GeV$^2$&13.15   GeV$^2$& 16 GeV$^2$\\ \hline
 $s=10^{-6}$ GeV$^2$&1.01 GeV$^2$& 1.46 GeV$^2$&2.5 GeV$^2$&4.4 GeV$^2$&7.9 GeV$^2$&11.2  GeV$^2$& 14.5 GeV$^2$& 17.6 GeV$^2$\\ \hline
 $s=10^{-7}$ GeV$^2$&1.08 GeV$^2$& 1.61 GeV$^2$&2.7 GeV$^2$&4.8 GeV$^2$& 8.65 GeV$^2$&12.3 GeV$^2$& 15.9 GeV$^2$& 19.4 GeV$^2$\\  \hline
 \end{tabular}
\caption{ The scale $k^2_{t1}$ ($50\%$ of the total cross section)
for longitudinal photons in
 DIS (CTEQ6)}
\end{center}
\end{table*}
\begin{table*}[htbp]
\begin{center}
\begin{tabular}{|l|l|l|l|l|l|l|l|l|}\hline
 &$Q^2=3$ GeV$^2$&$5$ GeV$^2$&$10$ GeV$^2$&$20$ GeV$^2$&$40$
 GeV$^2$&$60$ GeV$^2$&$80$ GeV$^2$&$100$ GeV$^2$
 \\  \hline
 $s=10^{-3}$ GeV$^2$&0.91 GeV$^2$& 1.07 GeV$^2$&1.35 GeV$^2$&1.78 GeV$^2$&2.3 GeV$^2$ &2.75 GeV$^2$&3.06 GeV$^2$&3.45 GeV$^2$\\ \hline
 $s=10^{-4}$ GeV$^2$&1.18 GeV$^2$& 1.42 GeV$^2$&1.82 GeV$^2$&2.42 GeV$^2$&3.25 GeV$^2$&3.9 GeV$^2$& 4.4 GeV$^2$& 5 GeV$^2$\\ \hline
 $s=10^{-5}$ GeV$^2$&1.37 GeV$^2$& 1.67 GeV$^2$&2.18 GeV$^2$&2.98 GeV$^2$&4.1 GeV$^2$& 5. GeV$^2$&5.8   GeV$^2$& 6.5 GeV$^2$\\ \hline
 $s=10^{-6}$ GeV$^2$&1.54 GeV$^2$& 1.91 GeV$^2$&2.57 GeV$^2$&3.57 GeV$^2$& 5. GeV$^2$&6.3 GeV$^2$& 7.3 GeV$^2$& 8.3 GeV$^2$\\  \hline
$s=10^{-7}$ GeV$^2$&1.76GeV$^2$& 2.21 GeV$^2$&2.95 GeV$^2$&4.3
GeV$^2$& 6.2 GeV$^2$&7.9 GeV$^2$& 9.35 GeV$^2$& 10.5 GeV$^2$\\
\hline

 \end{tabular}
\caption{ The scale $k^2_{t1}$ ($50\%$ of the total cross section)
for transverse photons in
 DIS (CTEQ6)}
\end{center}
\end{table*}

\begin{table*}[htbp]
\begin{center}
\begin{tabular}{|l|l|l|l|l|l|l|l|l|}\hline
 &$Q^2=3$ GeV$^2$&$5$ GeV$^2$&$10$ GeV$^2$&$20$ GeV$^2$&$40$
 GeV$^2$&$60$ GeV$^2$&$80$ GeV$^2$&$100$ GeV$^2$
 \\  \hline
 $x=10^{-2}$ GeV$^2$&38& 45 &50&56&60 &61&62.5&63\\ \hline
 $x=10^{-3}$ GeV$^2$& 61& 66 &71 &75&78 &80&80.5&81\\ \hline
 $x=10^{-4}$ GeV$^2$&73 &81 &84 &84.5&87 &88&88.5&89\\ \hline
 $x=10^{-5}$ GeV$^2$& 81&84  &87.5& 90 &92 &92.5&93&93.5\\ \hline
 $x=10^{-6}$ GeV$^2$& 87& 89.5 &92 &94&95 &95.5&96&96\\ \hline
 $x=10^{-7}$ GeV$^2$& 91& 93 &95 &96&97 &97.3&97.6&98\\ \hline
 \end{tabular}
\caption{ The pQCD contribution $(\%)$ in the total cross section,
that includes AJM contribution (CTEQ6)}
\end{center}
\end{table*}

\begin{table*}[htbp]
\begin{center}
\begin{tabular}{|l|l|l|l|l|l|l|l|}\hline
 &$Q^2=5$ GeV$^2$&$10$ GeV$^2$&$20$ GeV$^2$&$40$
 GeV$^2$&$60$ GeV$^2$&$80$ GeV$^2$&$100$ GeV$^2$
 \\  \hline
 $x=10^{-2}$ GeV$^2$& 1.93 &2. &2.08& 2.2&2.26  &2.3&2.34\\ \hline
 $x=10^{-3}$ GeV$^2$& 1.86 &1.94 &2.02& 2.1&2.13  &2.15&2.17\\ \hline
 $x=10^{-4}$ GeV$^2$& 1.74 &1.82 &1.88& 1.92&1.94  &1.95&1.97\\ \hline
 $x=10^{-5}$ GeV$^2$& 1.66  &1.72 &1.76&1.79 &1.8&1.81&1.82\\ \hline
 $x=10^{-6}$ GeV$^2$&1.6   & 1.63&1.67& 1.69&1.71&1.72&1.72\\ \hline
 $x=10^{-7}$ GeV$^2$&1.52  &1.56 &1.6&1.62 &1.64&1.66&1.66\\ \hline
 \end{tabular}
\caption{ The DVCS ratio R  (CTEQ6)}
\end{center}
\end{table*}

\begin{table*}[htbp]
\begin{center}
\begin{tabular}{|l|l|l|}\hline
 &$\Gamma =1$&$\Gamma=1/2$\\  \hline
 $s=10^4$ GeV$^2$& 1.6 GeV$^2$ &3.1 GeV$^2$\\ \hline
 $s=10^5$ GeV$^2$& 3 GeV$^2$   &5.6 GeV$^2$\\ \hline
 $s=10^6$ GeV$^2$& 6 GeV$^2$   &10 GeV$^2$\\ \hline
 $s=10^7$ GeV$^2$& 11 GeV$^2$  &21 GeV$^2$\\  \hline
 \end{tabular}
\caption{The scale $k^2_t$ for the onset of the black disk
 regime for the gluon dipole.}
\end{center}
\end{table*}

\begin{table*}[htbp]
\begin{center}
\begin{tabular}{|l|l|l|}\hline
 &$\Gamma =1$&$\Gamma=1/2$\\  \hline
 $s=10^4$ GeV$^2$&  - GeV$^2$  &1 GeV$^2$\\ \hline
 $s=10^5$ GeV$^2$&  - GeV $^2$  &2 GeV$^2$\\ \hline
 $s=10^6$ GeV$^2$&2 GeV$^2$  &5 GeV$^2$\\ \hline
 $s=10^7$ GeV$^2$&5 GeV$^2$  &10 GeV$^2$\\  \hline
 \end{tabular}
\caption{The scale $k^2_t$ for the onset of the black disk
 regime for the fermionic dipole.}
\end{center}
\end{table*}
\newpage
\clearpage
\begin{figure}[htbp]
\centerline{\epsfig{figure=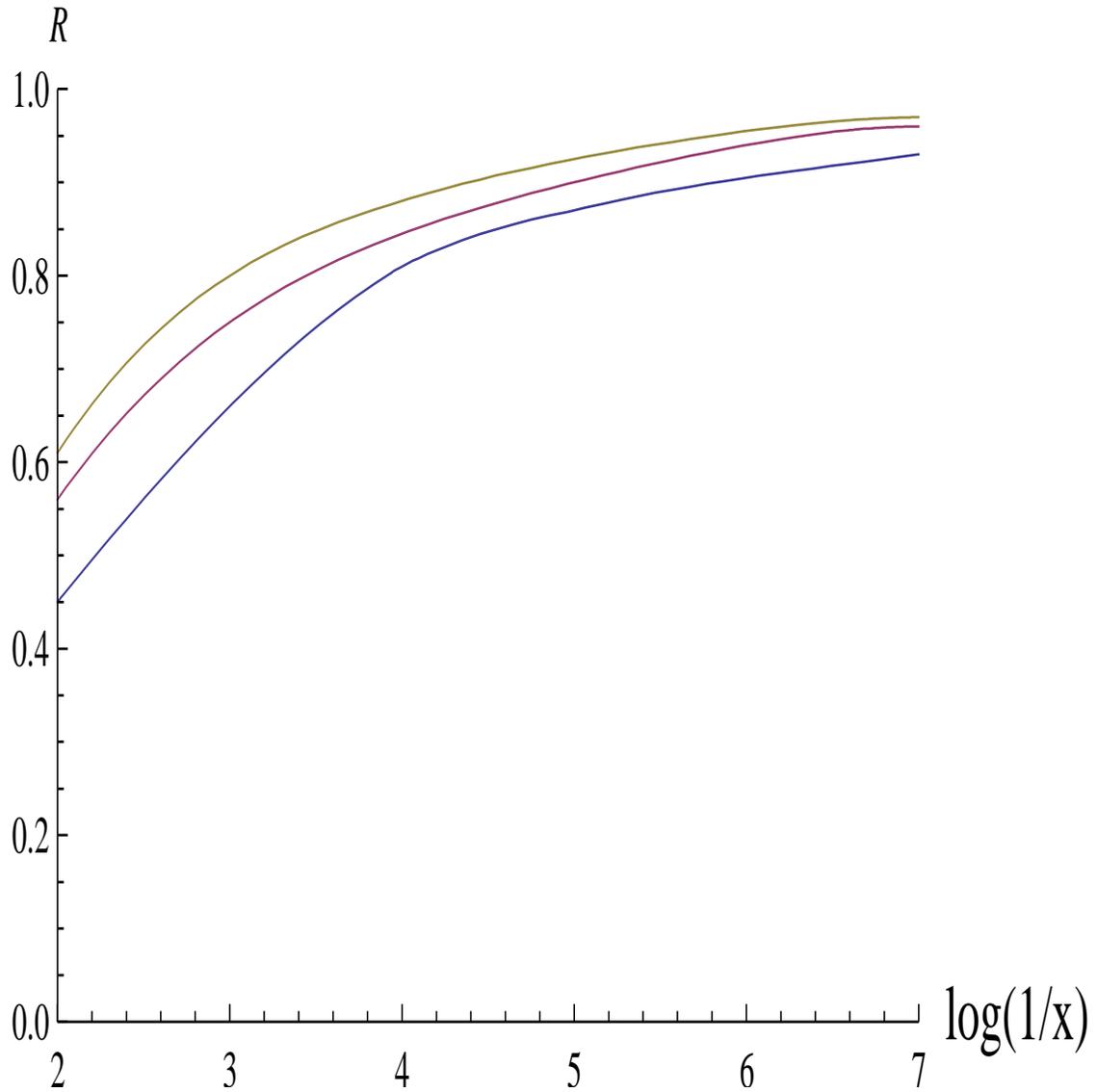,width=15cm,height=15cm,clip=}}
\caption{ The contribution R of pQCD to the total cross section,
that is a sum of pQCD and AJM model contributions. The cut off of
the AJM model is 0.35 GeV$^2$, for $Q^2=5$ GeV$^2$ (lower curve),
$20$ GeV$^2$ (middle curve), 40 GeV$^2$, the x axis corresponds to
$\log_{10}(1/x)$} \label{S7}
\end{figure}
\clearpage
\clearpage

\newpage\begin{thebibliography}{}
\bibitem{BFGMS} S. J. Brodsky, L. Frankfurt, J. F. Gunion, A. H. Mueller and M. Strikman, Phys. Rev. D50 (1994) 3134.
\bibitem{AFS}H. Abramowicz, L. Frankfurt and M. Strikman, Surveys
in High Energy Physics, 11 (1997) 51.
\bibitem{FMS} L. Frankfurt, G.A. Miller, M. Strikman, Ann. Rev.
Nucl. Part. Sci., 44 (1994) 501.
\bibitem{BBFS}B. Blaettel, G. Baym,
L. Frankfurt and M. Strikman, Phys. Rev. Lett., 70 (1993) 896.
\bibitem{Muellereikonal}  A.~H.~Mueller Nucl.\ Phys.\  B {\bf 415}, (1994) 373 .
\bibitem{CFS} J.C. Collins,  L. Frankfurt and M. Strikman, Phys. Rev. D56 (1997) 2982.
\bibitem{DGLAP} G. Altarelli and G.Parisi, Nucl. Phys., B126
(1977) 298; V.N. Gribov and L. N. Lipatov, Sov. J. of Nucl. Phys.,
15 (1972) 438,672; Yu.L. Dokshitser, Sov. Phys. JETP 46 (1977)
641.
 \bibitem{CH} S. Catani, M. Ciafaloni, F. Hauptmann, Nucl. Phys.
B366 (1991) 135.
\bibitem{CE} J. Collins, K. Ellis, Nucl. Phys., B360 (1991) 3.
\bibitem{Gribovspace-time} V. N. Gribov, Space-time description of hadron interactions at high-energies.
 In *Moscow 1 ITEP school, v.1 'Elementary particles'*,
65,1973. e-Print: hep-ph/0006158.
\bibitem{BFKL} E. Kuraev, V. Fadin, L. Lipatov, Sov. Phys.-JEP, 44
(1976) 443; 45 (1977) 199. I. Balitsky and L. Lipatov, Sov. J.
Nucl. Phys., 28 (1978) 822.
\bibitem{Sherman} L. Frankfurt and V. Sherman, Phys. Lett., 61B
(1976) 70.
 \bibitem{Rogers1}
T.~C.~Rogers, A.~M.~Stasto and M.~I.~Strikman, Unitarity
Constraints on Semi-hard Jet Production in Impact Parameter
Space,''arXiv:0801.0303 [hep-ph].
\bibitem{FGMS}
 V. Guzey, L. Frankfurt, M. Strikman, M.McDermott, Eur.J. of Physics, C16 (2000) 641.
\bibitem{FSW} L. Frankfurt, M, Strikman and C. Weiss,
Ann.Rev.Nucl.Part.Sci.55 (2005) 403-465.
\bibitem{Rogers2}
  T.~C.~Rogers and M.~I.~Strikman,
  Hadronic interactions of ultra-high energy photons with protons and  light  nuclei in the dipole picture,''
  J.\ Phys.\ G {\bf 32}, (2006) 2041.
\bibitem{Fadin} V.S. Fadin, hep-ph/9807528.
\bibitem{SP} F. Caporale, A. Papa, A. Sabio Vera, E. J. of
Physics,C53 (2008) 525.
\bibitem{FS} L. Frankfurt and M. Strikman, Phys. Rept., 160
(1988) 235.
\bibitem{FRS} L. Frankfurt, A.Radyushkin, M. Strikman, Phys. Rev.
D55(1997) 98.
\bibitem{Gribov} V. N. Gribov, Sov. Phys. JETP 30 (1969) 709.
\bibitem{Yennie}
T.H. Bauer, R.D. Spital, D.R. Yennie  , F.M. Pipkin,
Rev.Mod.Phys.50 (1978) 261, Erratum-ibid.51(1979) 407.
\bibitem{DDT} Yu. Dokshitzer,D. Diakonov and S. Troyan, Phys. Reports, 58  (1980) 269.
\bibitem{AS}M. Abramowitz and I. Stegun, Handbook of special
functions,Dover Publications, New York,1964.
\bibitem{Newman}Deep Inelastic Scattering at the TeV Energy Scale and the LHeC Project.
Paul Newman, (Birmingham U.) . Feb 2009. 13pp. To appear in the
proceedings of Ringberg Workshop on New Trends in HERA Physics
2008, Ringberg Castle, Tegernsee, Germany, 5-10 Oct 2008.
Published in Nucl.Phys.Proc.Suppl.191:307-319,2009. e-Print:
arXiv:0902.2292 [hep-ex]
\bibitem{ABF}G. Altarelli, S. Forte, R.D. Ball,
Nucl. Phys., B621 (2002) 359; B674 (2003) 459.
\bibitem{Ciafaloni}M. Ciafaloni, P. Colferai, G.P. Salam,
A. M. Stasto, Phys. Lett., 587 (2004) 87; Phys. Rev. D68 (2003)
114003.
\bibitem{GPD} L. Frankfurt, A. Freund, V. Guzey and M. Strikman,
Phys. Lett., B418 (1998) 345; Erratum-ibid, B429  (1998) 414.
\bibitem{FFS} L. Frankfurt, A. Freund and M. Strikman, Phys. Rev.
D58 (1998) 114001; Erratum D59 (1999) 119901.
\bibitem{FS2} L. Frankfurt and M. Strikman, Nucl. Phys., B316
(1989) 340.
  \bibitem{S} L. Schoeffel, Phys. Lett.,B658 (2007) 33.
\bibitem{LatestHera} A. Aktas et al, Eur. Phys. J., C48 (2006)
715; S. Chekhanov et al, arXiv:0812.2003 (hep-ex).
\bibitem{PRL}V. Guzey, L. Frankfurt, M. Strikman,M. McDermott, Phys.Rev.Lett., 87 (2001) 192301.
\bibitem{Frankfurt:2003td}
   L.~Frankfurt, M.~Strikman and C.~Weiss,
   Phys.\ Rev.\  D {\bf 69}, (2004) 114010.
\bibitem{GIP} L. B. Ioffe, V. Gribov, I. Pomeranchuk,
Sov. J. of Nucl. Phys., 2 (1966) 549.
\bibitem{Ioffe} B. L. Ioffe, Phys. Lett., B30 (1969) 123.
\bibitem{KS} Y. Kovchegov and M. Strikman, Phys. Lett., B516(2001) 314.
\bibitem{BFcoh} B.Blok and L.Frankfurt Phys.Lett.B630 (2005)
49-57.
\bibitem{BFS} B. Blok, L. Frankfurt, M. Strikman, ArXiv:0811.3737
(hep-ph)
\bibitem{Mueller02}
A.H. Mueller, Nucl.Phys.A702 (2003) 65-72.
\bibitem{Dumitru:2002qt}
  A.~Dumitru and J.~Jalilian-Marian,
  Phys.\ Rev.\ Lett.\  {\bf 89},(2002) 022301
  [arXiv:hep-ph/0204028].
\bibitem{DP}
  D.~Diakonov and V.~Y.~Petrov,
  Nucl.\ Phys.\  B {\bf 272} (1986) 457 .
\bibitem{FGS}
  L.~Frankfurt, V.~Guzey and M.~Strikman,
  Phys.\ Rev.\  D {\bf 71}, (2005) 054001.
\bibitem{Bohr} A. Bohr and B.R. Mottelson, Nuclear structure, v.1,
W.A. Benjamin, New York, 1969.
\bibitem{FSZ} L. Frankfurt , M. Strikman  , M. Zhalov
Phys.Lett.B616 (2005), 59-75.
\bibitem{Mark1} L. Frankfurt and M. Strikman, in Phys. Reports, 455 (2008) 105.
\bibitem{Gtrick} V.N.Gribov
 The theory of complex angular momenta: Gribov lectures on theoretical physics. Cambridge, UK: Univ. Pr.
 (2003).
\bibitem{CGC}L. McLerran and R. Venugopalan, Physical Review, D49 (1994),
2233,3352; D50 (1994) 2225; D59 (1999) 094002; see also L.
McLerran, Surveys High Energy Phys., 18 (2003) 101; Nucl. Phys.
A702 (2002) 49, for the latest reviews on the subject.
\bibitem{Mueller}
  A.~H.~Mueller Nucl.\ Phys.\  B {\bf 415}, (1994) 373.
\bibitem{H1} C. Adloff et al, Phys. Lett., B520 (2001) 183; S. Chekanov et
al, Nucl. Phys. B713 (2005) 3.
\end{thebibliography}
\end{document}